\documentclass[12pt,preprint]{aastex}

\shorttitle{Dynamical Simulations of NGC 2523 and NGC 4245}
\shortauthors{Treuthardt, Salo, \& Buta}

\begin{document}


\title{Dynamical Simulations of NGC 2523 and NGC 4245}


\author{P. Treuthardt\altaffilmark{1}, H. Salo\altaffilmark{2},
and R. Buta\altaffilmark{1}}\altaffiltext{1}
{Department of Physics and Astronomy, University of Alabama, Box 870324,
Tuscaloosa, AL 35487}\altaffiltext{2}{Division of Astronomy, Department of
Physical Sciences, University of Oulu, Oulu, FIN-90014, Finland}


\begin{abstract}

We present dynamical simulations of NGC 2523 and NGC 4245, two barred
galaxies (types SB(r)b and SB(r)0/a, respectively)
with prominent inner rings. Our goal is to estimate the bar
pattern speeds in these galaxies by matching a sticky-particle simulation
to the $B$-band morphology, using near-infrared $K_s$-band images to define
the gravitational potentials. We compare the pattern speeds derived
by this method with those derived in our previous paper
using the well-known Tremaine-Weinberg continuity equation method.
The inner rings in these galaxies, which are likely to be resonance features,
help to constrain the dynamical models. We find that both methods give the
same pattern speeds within the errors.

\end{abstract}




\section{Introduction}

The pattern speed, $\Omega_p$, is one of the most influential and elusive
parameters affecting the overall morphology of a galaxy. It remains
one of the fundamental unknowns of galaxy dynamics. A variety of
techniques for determining $\Omega_p$ have been published (e.g. Canzian 1993;
Buta \& Combes 1996; Puerari \& Dottori 1997; Salo et al. 1999; Weiner 
et al. 2001; Egusa et al. 2004; Zhang \& Buta 2007), though in recent years,
emphasis has been placed on an observational method developed by
Tremaine \& Weinberg (1984, hereafter TW). Studies of late that make
use of this method have focussed on estimating the
pattern speed of SB0 galaxies (e.g. Aguerri et al. 2003;
Corsini et al. 2003; Debattista et al. 2002).
These studies have
indicated that SB0 galaxies generally have "fast" bars, presumably
indicating halos of low central concentration.

The TW method is the most direct means of measuring the pattern speed of
a galaxy. Here, the pattern speed of a bar can be estimated from the
luminosity weighted mean
line-of-sight velocities, $\langle$$V$$\rangle$, and luminosity weighted mean
positions, $\langle$$X$$\rangle$, (both relative to the central values)
of a tracer that obeys the
continuity equation. These quantities are to be measured along lines
parallel to a barred galaxy's projected disk major axis and are related as 
follows,
\begin{equation}\Omega_p \sin{i} = \frac{\langle V \rangle}{\langle X \rangle}\end{equation}
where $i$ is the galaxy's inclination.

One of the great advantages of the TW method is that it provides pattern speeds
independent of any assumptions concerning features traced by the pattern
speed. For example, the TW method can used to directly test the ideas that
bars end near their own corotation resonances, and that rings are associated
with specific resonances. However, this advantage is somewhat offset
by the limited scope of applicability of the method. The most significant
hindrance is that the bar major axis must lie
at approximately 45$^{\circ}$ to the major axis of the projected disk
in order to obtain
the maximum TW signature strength, with angles between about 20$^{\circ}$
and 70$^{\circ}$ considered suitable (Gerssen et al. 2003).
The galaxy itself
must have a preferred inclination between 50$^{\circ}$ and 60$^{\circ}$
(Debattista 2003) in order to avoid large
uncertainties. Another problem is identifying a pattern speed tracer
with a strong spectral signature that obeys the continuity equation. This
means that the tracer must
be something that is neither created nor destroyed and significant star
formation
would violate the equation. This is the reason that applications to spirals
have been
more limited. Hernandez et al. (2005) discuss the application of the TW method
to atomic, molecular, and ionized gas phases in spiral galaxies.
SB0 galaxies were the first objects to which the method was
applied, using stellar absorption lines (e.g., Merrifield \& Kuijken 1995;
Gerssen et al. 1999). These galaxies can
have strong bar patterns but lack the dust and star formation that would
complicate
similar measurements for later-type systems. Recently though, a TW test of
an n-body model by
Gerssen \& Debattista
(2007) indicated that dust absorption effects may be partially mitigated by
star formation on the leading edges of a bar, creating a minimal impact on the
derived $\Omega_p$.
Avoiding these potentially problematic factors
significantly diminishes the number of galaxies to which this method can be
applied.

A less direct, yet more widely applicable method of determining $\Omega_p$ in
galaxies is through simulation modeling (e.g., Salo et al. 1999; Rautiainen,
Salo, \& Laurikainen 2005). This involves modeling the response of gas in
a rigid galaxy potential, where the gas is modeled in terms of dissipatively
colliding, or "sticky," particles. These sticky particles simulate the cold gas 
component of a galaxy, where regions of star formation occur due to cloud-cloud
collisions (Levinson \& Roberts 1981). The 
gravitational potential is derived from a near-IR image. Different dynamical
parameters are varied until the morphology and kinematics of the numerically
simulated galaxy visually matches, as closely as possible, the blue light
morphology. The morphology of resonance rings is sensitive to the 
underlying pattern speed of a galaxy and serve as a good $\Omega_p$ estimator. 
We have deep $B$-band images with good resolution in order to 
compare the observed low velocity dispersion components to our models. Since 
resonance rings generally form in the gas component and
show emission from young stars, the observed $B$-band 
morphology is best suited for this purpose. Sticky particle modeling is 
similar to hydrodynamic modeling, such as SPH (smoothed particle 
hydrodynamics), only the particles are not subjected to
non-gravitational forces. Therefore, the motions of particles are not affected
by the positions of other particles (i.e through pressure gradients), except
through dissipation in physical impacts. This difference means that SPH is 
better suited for modeling the warm
gas phase of a galaxy, while the sticky particle method is better suited for 
approximating the cold gas phase (Merlin \& Chiosi 2007).
In Salo et al. (1999), which addresses IC 4214, it was estimated that sticky
particle
modeling can yield $\Omega_p$ within about 10-20\%, even when allowing for
uncertainties in the bar amplitude (due to unknown vertical extent, unknown
dark halo contribution, and possible variations in the M/L ratio) and the
numerical parameters involved in the sticky particle method itself. Direct 
comparisons between pattern speeds derived from sticky particle and 
SPH methods of modeling are not published, although a comparison of results 
from both methods is 
discussed by Bournaud et al. (2005) in regards to the gravity torques between 
a stellar bar and gas. It was found that the tree-SPH code by Semelin \& Combes
(2002) produced the same order value as in sticky particle simulations and 
observations.

In this paper, we compare the kinematic bar pattern speeds of
NGC 2523 and NGC 4245, measured through the use
of the TW method (Treuthardt et al. 2007), to the dynamical bar pattern speeds
derived through simulations. In particular we base our simulation estimates
solely on morphological comparisons and check whether $\Omega_p$ is consistent
with direct measurements. If so, this would justify the application of the
simulation method to a large number of galaxies observed in various near-IR
surveys, not necessarily possessing such extensive kinematical data as IC 4214.
NGC 2523 is an SB(r)b galaxy with a closed inner ring and multiple arms (see
Figure 1).
A color index map (Buta, Corwin, \& Odewahn 2007) shows that the spiral
arms and part of the inner ring are sites of recent star formation.
NGC 4245 is an SB(r)0/a galaxy with strong nuclear and inner resonance ring
features that are also prominent blue features
in a Buta, Corwin, \& Odewahn (2007)
color index map. NGC 4245 is an excellent candidate for such a test due to
the good statistical fit of the measured pattern speed. The
resonance features apparent in these galaxies help to constrain the
dynamical models used to determine the bar pattern speed because the sizes
and shapes of the rings are sensitive to the pattern speed. Since
gas velocity maps of these galaxies do not exist, we are not able to compare
the modeled gas kinematics to observations.

\section{Analysis}
\subsection{Determining the Potential}
In order to simulate the dynamics of the galaxies, a gravitational
potential was derived by assuming that the
near-infrared light distribution effectively traces the mass
distribution. This assumption is justified since the amount of dark matter
within the $R_{25}$ optical disk radius is small compared to the amount of
visible matter, allowing maximal disk models to account for the observed
rotation (see Quillen, Frogel, \& Gonzalez 1994 and references therein). The
potential may then be derived from the convolution
of the density with the function 1/$R$ (Quillen, Frogel, \& Gonzalez
1994), combined with an assumption for the vertical density
distribution. For NGC 2523 and NGC 4245, the potentials were derived from
2.15$\mu$m $K_{s}$-band images obtained for the Near-Infrared S0 Survey
(Laurikainen, Salo, \& Buta 2005, Buta et al. 2006, Laurikainen et al. 2006)
in January 2003 with the 2.5m Nordic Optical Telescope (NOT) and in May
2006 with the 4.2m William-Herschel Telescope (WHT),
respectively. During the NOT run, $B$ and $V$-band images were also taken to
help define the orientation parameters of the two galaxies. These
values are summarized by Treuthardt et al. (2007).

The $K_{s}$-band light distribution is useful for estimating the potential
due to the reduced extinction effects from dust and the
weakened influence of young Population I complexes. Before the
gravitational potential was derived, the $K_{s}$-band images were
converted into a surface mass density by assuming a mass-to-light
($M/L$) ratio accounting for the average disk $V$-$K_{s}$ color gradient using
formulae from Bell \& de Jong (2001). The ``formation epoch with bursts'' $M/L$
model listed in Table 1 of Bell \& de Jong (2001) was applied
to the
azimuthally averaged $K_{s}$-band images, which were then used to derive a
gravitational potential. These color profiles are shown in Figure 2.
Note that although it would be ideal to apply $M/L$ corrections on a pixel
by pixel basis, instead of an azimuthal average, the much larger uncertainties
on pixel colors makes this less practical. Thus, we are only correcting the
axisymmetric backgrounds of the galaxies for stellar $M/L$ effects.

The gravitational potentials were calculated
by first using an iterative two-dimensional
bulge-disk-bar decomposition identical to that described by
Laurikainen et al. (2004). This method models the bulge in terms of a
S\'{e}rsic $r^{1/n}$ function, the disk in terms of an exponential,
and the bar in terms of a Ferrers function. A more sophisticated approach
is used by Laurikainen et al. (2006). The parameters are iterated on a
sky-subtracted image cleaned of foreground and background objects.
The
bulge component was removed from the $K_{s}$-band light
distribution and the disk was then deprojected to a face-on orientation.
The disk light distribution was approximated by a Fourier
decomposition and the disk gravity was calculated using the even
components from $m=0$ to $m=8$, as was done by Rautiainen et al.
(2005). The odd Fourier components of the disk are small compared to the
even components. For example, the maximum relative tangential perturbation
$F_T/F_R$ is 0.56 for NGC 2523 when only the even components are taken 
into account. $F_T/F_R$ increases by approximately 0.02 to 0.58 when the 
odd Fourier components from $m=1$ to $m=9$ are included. Here $F_T$ stands 
for the amplitude of the tangential force, while $F_R$ is the azimuthally
averaged radial force.
Typical vertical scale heights of 1/5 and 1/4
the radial scale length were assumed for the disk components of NGC 2523 and
NGC 4245, respectively. These scale heights were selected based on the
empirical relation of scale height to radial scale length of a galaxy discussed
by de Grijs (1998). Overall,the effect of scale height on $F_T/F_R$ is not 
large. For example, if the
ratio of vertical scale height to radial scale length of NGC 2523 is changed
from 1/5 to 1/10, the maximum value of $F_T/F_R$ would change from 0.56 to
0.70. The gravitational potential of
the bulge was also added to the disk potential under the assumption
that the bulge mass is spherically distributed. A constant $M/L$
scale factor was applied to each potential so that the derived circular-speed
curve reached the same maximum velocity used by
Treuthardt et al. (2007; see Figures 3 and 4). No gas rotation curves of these
galaxies exist for comparison with the modeled circular-speed curves.


\subsection{Simulations}

The simulation code used to model NGC 2523 and NGC 4245 was written by
H. Salo,
the details of which can be found in Salo et al.
(1999; see also Salo 1991 for the treatment of particle impacts). The behavior 
of a two-dimensional disk of 20,000
inelastically colliding gas particles and 100,000 non-colliding stellar
test particles was simulated in the
determined potential. The initial particle distribution was exponential and the
particle velocities were calculated with the epicyclic approximation, using a 
radial velocity dispersion amounting to 10\% of the circular velocity at each
radius. The circular velocity was calculated using the total axisymmetric 
component of the gravity potential. The non-axisymmetric potential component of
the bar potential was turned on gradually and
reached full strength at two bar rotations. The main parameter that was
varied was the bar
pattern speed. The morphology of the simulated gas distribution was
compared to the morphology observed in the $B$-band image of each
galaxy (Treuthardt et al. 2007). We specifically compared the size of the
inner ring and the overall structure outside of this ring.

In an effort to increase the signal-to-noise of the simulation snapshots at 
the bar
rotation period of interest, we aligned and coadded particle snapshots
within $\pm$0.2 bar
rotation periods in 0.1 bar rotation period increments. The net result is a
snapshot of 100,000 gas and 500,000 star particles at a specified bar
rotation period. This
can be done because after the initial transient evolution
related to turn-on of the bar, the morphology of the simulated galaxies
does not change significantly on such a relatively small timescale. In 
practice, the results were indistinguishable from using a five-fold number
of actual particles of appropriately reduced size to keep the impact 
frequency constant.

The pattern speed of the simulated galaxies was varied in such a way
that the ratio of the corotation radius ($R_{CR}$) to the bar radius
($R_{B}$) ranged from 2.0 to 1.0 in increments of 0.1 (see Table 1). The
deprojected
bar radius was estimated to be 33.5$\arcsec$ for NGC 2523 and 38.1$\arcsec$
for NGC 4245 (Treuthardt et al. 2007).
In Figures 3 and 4 (upper right), we plot the Lindblad precession
frequency curves
derived from the simulated circular-speed curves estimated from the
$K_{s}$-band images of the galaxies. These curves show how resonance
locations vary with angular velocity in the linear (epicyclic)
approximation.

It should be
noted that these simulations assume a single pattern speed is present,
when in fact many galaxies may have multiple pattern speeds. Pfenniger \&
Norman (1990) suggested that nuclear bars represent independent instabilities
that could have their own pattern speeds. There is evidence for this in the
random alignments observed between primary and secondary bars (e.g. Buta
\& Crocker 1993; Wozniak et al. 1995; Friedli et al. 1996) and even from
direct observations (e.g. Corsini et al. 2003).
Sellwood \& Sparke (1988) showed that in their
n-body simulations of naturally forming bars and spirals, the spirals tend
to have a lower pattern speed than the bar.
More recently, Rautiainen \& Salo (1999, 2000) examined the effect of
multiple pattern speeds in barred galaxies over a Hubble time and with respect
to the ring formation. Their simulations demonstrated that the presence of
several patterns at the region of the outer Lindblad resonance
(OLR; $\Omega_p=\Omega+\kappa/2$, where $\Omega$ is the circular angular
velocity and $\kappa$ is the epicyclic frequency)
does not inhibit the formation of an outer ring, but can cause cyclic
changes in its morphology (Fig. 10 in Rautiainen \& Salo 2000). However, even
when slower modes are present and affect the shape of the outer ring, the
ring seems to be related to the OLR of the bar, that is the ring size
accomodates the change in OLR distance due to bar slow-down.
Nevertheless, we conclude that the possiblilty of multiple pattern
speeds is not
a serious limitation on our study because our focus is on the bars and
morphology immediately surrounding and outside the bars.

\section{Results and Discussion}


In order to estimate $\Omega_p$ for these galaxies, we would like to find
the $\Omega_p$ value that best visually matches the simulated and observed
morphologies. Because of their smaller velocity dispersion compared to stars, 
gas morphology is more sensitive to $\Omega_p$. Since $B$-band images tend
to show
details of gas morphology, our visual comparisons will be between models
and $B$-band images.
Note that factors other than $\Omega_p$ affect the morphology of a
simulated galaxy. These factors include the time when the features
are examined,
impact frequency from the assumed cross section of the particles,
contribution of a possible dark halo component, and the strength of
$F_T/F_R$ via the assumed vertical profile.

Figures 5 and 6 show the effect of different bar pattern speeds (or
$R_{CR}/R_{B}$ values), at an equal
number of bar rotation periods, on the
morphologies of NGC 2523 and 4245, respectively. In the case of NGC 2523,
larger values of $\Omega_p$ (corresponding to smaller values of $R_{CR}/R_{B}$;
see Table 1) cause the inner ring to become smaller and less
oval until it is no longer able to form. When $R_{CR}/R_{B}$ = 1.4, a
gaseous bar is clearly apparent, the inner ring is approximately
the same size as seen in the $B$-band, and the simulated morphology
outside of the inner ring resembles that of the observed galaxy.
In the case of NGC 4245, larger values of $\Omega_p$ also cause
the inner ring to become smaller until it is no longer able to form.
When $R_{CR}/R_{B}$ is 1.6 or greater, so called "banana" orbits are
apparent, while when $R_{CR}/R_{B}$ is 1.4 or less, the $L_4$ and
$L_5$ Lagrangian point regions are nearly cleared of gas except for
some small collections of particles.

The evolution of the simulated morphologies of NGC 2523 and NGC 4245
is shown in Figure 7. Here a single $\Omega_p$
is examined at a different
number of bar rotation periods. In the case of NGC 2523, the gas
morphology of the simulation appears settled after about 4 bar
rotations while the sharp features appear to slowly diffuse.
In the case of NGC 4245, the gas morphology appears
settled after about 8 bar rotations, but the $L_4$ and $L_5$ regions
continue to be cleared of particles. As an aside, we changed the 
number of simulation steps per orbit in order to examine the effects 
on the models of NGC 2523. We found that changing our standard of 
approximately 350 steps per orbit at the bar radius (33.5$\arcsec$) 
to 700 steps per orbit did not produce a noticably different morphology.

The effect of the particle impact frequency on the simulated morphology
of the galaxies is shown in Figure 8. Here we see how the morphology changes 
when the gas particle radius of 1$\times$10$^{-2}$
arcseconds is changed by a factor
of 1/10, 1, and 10. As the particle cross section increases, the impact
frequency also increases and the gaseous features become sharper and
well-defined. For NGC 2523, the similarity in morphology between the 
observation and the model, which uses our nominal particle cross section, 
is not as apparent 
when the cross section is
changed by an order of magnitude. The typical impact freqencies of the 
inner and outer ring regions of this galaxy, using our nominal particle
cross section, are 1.5 and 0.8 
impacts/particle/bar rotation, respectively. For NGC 4245, the typical impact
frequencies of the inner and outer ring regions are 1.4 and 0.4 
impacts/particle/bar rotation, respectively. When the impact freqencies are 
much less than 1 impact/particle/bar rotation the morphology approaches that 
of non-colliding test particles. This is evident when comparing the morphology 
of the galaxies using 1/10 the nominal particle radius in Figure 8 to the 
stellar test particle morphology in Figure 13. If the initial distribution 
of particles is uniform instead of exponential, the impact frequencies would
increase in the outer parts of the galaxy relative to the inner parts.

The simulations shown thus far do not
include a dark halo component. The inclusion of a halo also has
implications in the morphology of the simulations. In Figures 9 we show
the effect of a halo component, based on the universal rotation curve of 
Persic et al. (1996), on the circular-speed curve and Lindblad
precession frequency curves of NGC 2523 and NGC 4245. The added halos
were assumed to be isothermal spheres with core radii of 147$\arcsec$ 
and 83$\arcsec$ and asymptotic velocities of 159 km s$^{-1}$ and
137 km s$^{-1}$ for NGC 2523 and NGC 4245, respectively. The mass of the 
halo component of NGC 4245 was scaled by approximately 1/3, 
with respect to the universal rotation curve model, in order to prevent the 
total circular-speed curve from rising significantly at large radii 
(r $>$ 60$\arcsec$). A constant M/L
scale factor was again applied to each potential so that the derived
rotation curve reached the same maximum velocity used by Treuthardt
et al. (2007). In Figure 10, the effects of the halo component are shown for
$R_{CR}/R_{B}$ = 1.4 and 1.5 for NGC 2523 and NGC 4245, respectively.
In the case of NGC 2523, the halo contribution has no significant effect on the
morphology of the galaxy. In the case of
NGC 4245, there is significant structure outside the inner ring. Gas particles 
appear to collect into prominent "banana" orbits. The 
reduced perturbation does not lead to the clearing of $L_4$ and $L_5$ as seen 
in the models without a halo.

We also examine the effects of varying the bar amplitude, without adding 
a halo, on the modeled gas morphology of the galaxies in Figures 11 and 12. 
In these models, the
non-axisymmetric Fourier components of the potential were multiplied by a
factor while the axisymmetric component was kept intact. We examine models
using our best estimate of $\Omega_p$ found in this paper and the average
$\Omega_p$ estimated by Treuthardt et al. (2007). The bar amplitude is varied
from a factor of 1.25 to 0.25. These models serve to describe the possible 
effects of
uncertainties in elements such as the bar height and halo contribution. 
A comparison of the total circular-speed curves of NGC 4245 in Figures 4 and 
9 suggests that the total radial force increases by approximately 1.4 times at 
80$\arcsec$ with the additon of our halo component. This corresponds to a 
reduction in $F_T/F_R$ by a factor of approximately 0.71. Therefore, our 
models created using different bar amplitudes explore a wider range of 
uncertainties in the halo contribution than in Figure 10.

In the case of NGC 2523 using the best estimate of $\Omega_p$ found in this 
paper, an inner ring does not form when a bar amplitude 
factor of 1.25 is applied. A factor of 0.75 produces an inner ring that 
appears somewhat stretched along the galaxy major axis. A gas bar is not
produced and the outer spiral arms are more clearly defined compared to 
the model using a factor of 1.00. Models produced using bar amplitude factors 
of less than 0.75 have morphologies that are clearly dissimilar to what is 
observed. Using the average $\Omega_p$ estimated by Treuthardt et al. 
(2007) and different bar amplitude factors, the models do not display 
morphology similar to what is observed in the $B$-band. Most significantly, 
the inner ring shape does not match the observations in any case. Based on 
these results, it appears that our best estimate of $\Omega_p$, along with a 
bar slightly weaker than our nominal value, recreates the observed morphology 
most accurately. 

Using the best estimate of $\Omega_p$ found in this 
paper for NGC 4245, we find that a bar amplitude factor of 1.25 produces an 
inner ring that is thinner and slightly more elliptical than what is seen in 
both our nominal case and the $B$-band image.
Models with a bar amplitude factor of less than 1.00 produce thicker inner 
rings and show banana orbits, along with other structure, outside the inner 
ring region that is not observed in the galaxy. Models produced using the    
average $\Omega_p$ estimated by Treuthardt et al. (2007) and different bar 
amplitude factors have inner ring morphologies that are clearly dissimilar to 
the observations. Although the error limits of this observationally derived 
$\Omega_p$ estimate are large, this could be an indication of multiple pattern 
speeds. The Tremaine-Weinberg method is more likely to measure the inner-most 
pattern speed since the signal-to-noise ratio is higher. This 
leads us to believe that our model with the best estimate of $\Omega_p$ and 
nominal bar amplitude are correct.

For NGC 2523, good agreement between the observed and simulated size of the
inner ring and morphology outside this ring is
obtained with
$R_{CR}/R_{B}$ = 1.4 $\pm$ 0.1 (see Figures 5 and 13). This corresponds to a
pattern speed
of 6.2$^{+0.5}_{-0.5}$ km s$^{-1}$ arcsec$^{-1}$, or 25.0$^{+2.0}_{-2.0}$
km s$^{-1}$ kpc$^{-1}$ with an assumed
distance of 51.0 Mpc (Kamphuis et al. 1996). For NGC 4245,
agreement is obtained with $R_{CR}/R_{B}$ = 1.5 $\pm$ 0.1 (see Figures
6 and 13).
This corresponds to a pattern speed of 2.7$^{+0.3}_{-0.2}$ km s$^{-1}$
arcsec$^{-1}$, or 43.5$^{+4.9}_{-3.3}$ km s$^{-1}$ kpc$^{-1}$ with an assumed
distance of 12.6 Mpc (Garc\'{i}a-Barreto et al. 1994). These $\Omega_p$ values
agree within the errors, though marginally, with the values determined by
Treuthardt et al. (2007) through the completely independent TW method
(see Table 2). In Table 3
we list the inferred resonance locations for our best fitting models. We also
compare the observed and modeled ring major axis radii and axis ratios.

Our best
simulations do not match the observed morphologies exactly. The
simulated inner region of NGC 2523 shows gas patricles collecting near the
ends of the bar, close to the inner ring, which could be interpreted as
regions of star formation (see Figure 13). There is no such star formation
evident in the
observed bar of NGC 2523. The structure outside the observed inner ring of
NGC 2523 is complicated and slightly asymmetric, whereas the simulated
morphology is necessarly bisymmetric since only even $m$ values were used to
calculate the disk gravity. A comparison of the simulated stellar 
and $K_s$-band morphology shows even less of a similarity. 
The main families of periodic stellar orbits, $x_1$ and $x_2$, are easily 
recognized in the simulated 
morphology of the galaxy but are indiscernable in the observed stellar image.
In general, the bar supporting $x_1$ family of orbits will not extend beyond 
the corotation radius of a galaxy, which places an upper limit on the bar 
pattern speed. For NGC 2523, these orbits appear to extend the 
length of the bar, which is expected.

The obvious difference
between our best simulation and the observed $B$-band morphology of NGC 4245 
is within
the inner ring. The simulation produces a nuclear ring approximately 3.3
times larger than what
is seen in the $B$-band (see Table 3 and Figure 13). A faster pattern speed
produces a
nuclear ring more similar in size to the observed ring, but the external
structures are vastly
different. It is possible that the observed morphology is due to more than one
pattern speed (Rautiainen et al. 2005), which is not accounted for in our
models. Additionally, outside the simulated
inner ring is a diffuse outer ring that is not obvious in the observed image.
Even our models that include a halo show an outer ring (see Figure 10). 
It was noted by Treuthardt et al. (2007) that NGC
4245 is one of the more H I-deficient galaxies in the Coma I group and that
gas stripping might explain the lack of an observed outer ring. Our models
do not account for such an interaction with the galaxy's environment. 
The simulated stellar 
and $K_s$-band morphology of this galaxy appear nearly as dissimilar as they 
did for NGC 2523. Both the stellar $x_1$ and $x_2$ orbits are recognizable in 
the simulated morphology, but are likewise indiscernable in the $K_s$-band 
image. The bar supporting $x_1$ family of orbits appear to extend the 
length of the bar. The simulated stellar component also displays faint banana 
orbits that are not seen in the observations. However, these comparisons to 
non-colliding test particles should not be taken too literally, as no attempt
was made to match their velocity distributions with observations. Rather,
they serve to illustrate the extent of various orbital families.

\section{Conclusions}

Using near-IR images to trace the stellar disk potential, we
have estimated the pattern speeds and resonance locations of
NGC 2523 and NGC 4245 using the
numerical simulation method of Salo et al. (1999; also
Rautiainen, Salo, \& Laurikainen 2005). This is  
completely independent of the approach used by Treuthardt et al. (2007).
We have examined the effects of varying the bar pattern speed 
(or $R_{CR}/R_{B}$), number of bar rotation periods, 
number of simulation steps per orbit, particle cross section, 
halo contribution, and bar amplitude on the simulated galaxy morphology. Our 
best models were chosen based on the closest match between the observed 
$B$-band and modeled morphology. In particular, we compared the size of the 
inner ring and the morphology outside of the ring. Our best simulation model 
provides
an interpretation of NGC 2523 which places corotation at 1.4 $\pm$ 0.1 times
the estimated bar radius of 33.5$\arcsec$, consistent with the results of
Treuthardt et al. (2007). For NGC 4245, a similar model
gives corotation at 1.5 $\pm$ 0.1 times the estimated bar
radius of 38.1$^{\arcsec}$, which marginally agrees with the results of
Treuthardt et al. (2007), within the errors. The values we derived through 
simulation modeling are robust against the various parameters examined.

Dynamical simulations are of great value as a means of estimating pattern
speeds because they are more widely applicable than the TW method. Galaxies
where the TW method is not appropriate, due to unfavorable inclinations or
bar-disk position angles, can still be simulated to determine $\Omega_p$.
Thus, as further tests like ours are made, reliable pattern speeds will
become available for larger numbers of galaxies.

P. Treuthardt and R. Buta acknowledge the support of NSF Grants AST-0205143
and AST-0507140 to the University of Alabama. H. Salo acknowledges the
support of the Academy of Finland. We thank E. Laurikainen and J. H. Knapen
for the use of the NIRS0S $K_s$-band images that we used for our simulations.

\clearpage


\clearpage

\begin{deluxetable}{lcccc}
\tabletypesize{\scriptsize}
\tablewidth{0pc}
\tablecaption{Simulation Data}
\tablehead{
\colhead{Simulation}
&\colhead{$R_{CR}/R_{B}$}
&\colhead{$\Omega_p$ (km s$^{-1}$ arcsec$^{-1}$)}
&\colhead{$\Omega_p$ (km s$^{-1}$ kpc$^{-1}$)\tablenotemark{a}} 
}

\startdata
NGC 2523 without halo	& 2.0  &  4.1 &  16.5  \cr
          		& 1.9  &  4.3 &  17.6  \cr
	      	    	& 1.8  &  4.6 &  18.8  \cr
      		        & 1.7  &  5.0 &  20.0  \cr
	          	& 1.6  &  5.3 &  21.5  \cr
			& 1.5  &  5.7 &  23.1  \cr
			& 1.4  &  6.2 &  25.0  \cr
			& 1.3  &  6.7 &  27.1  \cr
			& 1.2  &  7.3 &  29.5  \cr
			& 1.1  &  8.0 &  32.2  \cr
			& 1.0  &  8.6 &  34.9  \cr
NGC 2523 with halo      & 1.4  &  6.2 &  25.1  \cr
NGC 4245 without halo	& 2.0  &  1.8 &  29.9  \cr
          		& 1.9  &  1.9 &  31.9  \cr
	      	    	& 1.8  &  2.1 &  34.2  \cr
      		        & 1.7  &  2.2 &  36.8  \cr
	          	& 1.6  &  2.4 &  39.9  \cr
			& 1.5  &  2.7 &  43.5  \cr
			& 1.4  &  2.9 &  47.7  \cr
			& 1.3  &  3.2 &  52.7  \cr
			& 1.2  &  3.6 &  58.4  \cr
			& 1.1  &  4.0 &  64.8  \cr
			& 1.0  &  4.4 &  71.5  \cr
NGC 4245 with halo      & 1.5  &  2.9 &  51.5  \cr
\enddata
\tablenotetext{a}{with an assumed distance of 51.0 Mpc for NGC 2523 and 12.6 Mpc 
for NGC 4245}
\end{deluxetable}

\clearpage

\begin{deluxetable}{lccc}
\tabletypesize{\scriptsize}
\tablewidth{0pc}
\tablecaption{Pattern Speed Estimates}
\tablehead{
\colhead{Galaxy} 
&\colhead{$\Omega_p$\tablenotemark{a}        \tablenotemark{b}}
&\colhead{$\Omega_p$\tablenotemark{a}}\\ 
\colhead{}
&\colhead{(Treuthardt et al. 2007)}
&\colhead{(Dynamical Simulation)}
}

\startdata
NGC 2523  & 6.6 $\pm$ 1.6  &  6.2$^{+0.5}_{-0.5}$ \cr\\ 
NGC 4245  & 4.7 $\pm$ 1.9  &  2.7$^{+0.3}_{-0.2}$ \cr\enddata 
\tablenotetext{a}{ km s$^{-1}$ arcsec$^{-1}$}\tablenotetext{b}{ Inclinations 
of 49.7$^{\circ}$ for NGC 2523 and 35.4$^{\circ}$ for NGC 4245 were assumed
in these values.}
\end{deluxetable}

\clearpage

\begin{deluxetable}{llccccccc}
\tabletypesize{\scriptsize}
\tablewidth{0pc}
\tablecaption{Galaxy Resonance and Ring Properties\tablenotemark{a}}
\tablehead{
\colhead{Galaxy} 
&\colhead{}
&\colhead{ILR}
&\colhead{CR}
&\colhead{OLR}
&\colhead{$a_{n}$}
&\colhead{$(b/a)_n$}
&\colhead{$a_{i}$}
&\colhead{$(b/a)_i$}\\
\colhead{(1)}
&\colhead{}
&\colhead{(2)}
&\colhead{(3)}
&\colhead{(4)}
&\colhead{(5)}
&\colhead{(6)}
&\colhead{(7)}
&\colhead{(8)}
}

\startdata
NGC 2523 & Model\tablenotemark{b}       & 14.5$\arcsec$ & 47.0$\arcsec$ & 69.5$\arcsec$ & n/a                           & n/a             & 38$\arcsec$ $\pm$ 4$\arcsec$ & 0.66 $\pm$ 0.13  \\
         & Observation\tablenotemark{c} &               &               &               & n/a          & n/a  & 35.3$\arcsec$                & 0.74           \cr\\
NGC 4245 & Model\tablenotemark{d}       & 23.5$\arcsec$ & 57.2$\arcsec$ & 82.0$\arcsec$ & 16$\arcsec$ $\pm$ 1$\arcsec$  & 0.75 $\pm$ 0.08 & 46$\arcsec$ $\pm$ 5$\arcsec$ & 0.61 $\pm$ 0.13  \\ 
         & Observation\tablenotemark{c} &               &               &               & 4.8$\arcsec$ & 0.92 & 40.6$\arcsec$                & 0.77           \\\enddata
\tablenotetext{a}{Explanation of columns: (1) Galaxy name; (2) inner Lindblad resonance radius; (3) corotation radius; (4) outer Lindblad resonance radius; (5) nuclear ring major axis radius; (6) nuclear ring axis ratio; (7) inner ring major axis radius; (8) inner ring axis ratio. All of the values are in the disk-plane of the galaxy.}
\tablenotetext{b}{Best fitting model with no halo, $R_{CR}/R_{B}$ = 1.4, and at 7 bar rotations.}
\tablenotetext{c}{Parameters from Treuthardt et al. (2007).}
\tablenotetext{d}{Best fitting model with no halo, $R_{CR}/R_{B}$ = 1.5, and at 12 bar rotations.}
\end{deluxetable}



\clearpage

\begin{figure}
\figurenum{1}
\includegraphics[angle=0,trim=0 75 0 0,clip=true,scale=0.75]{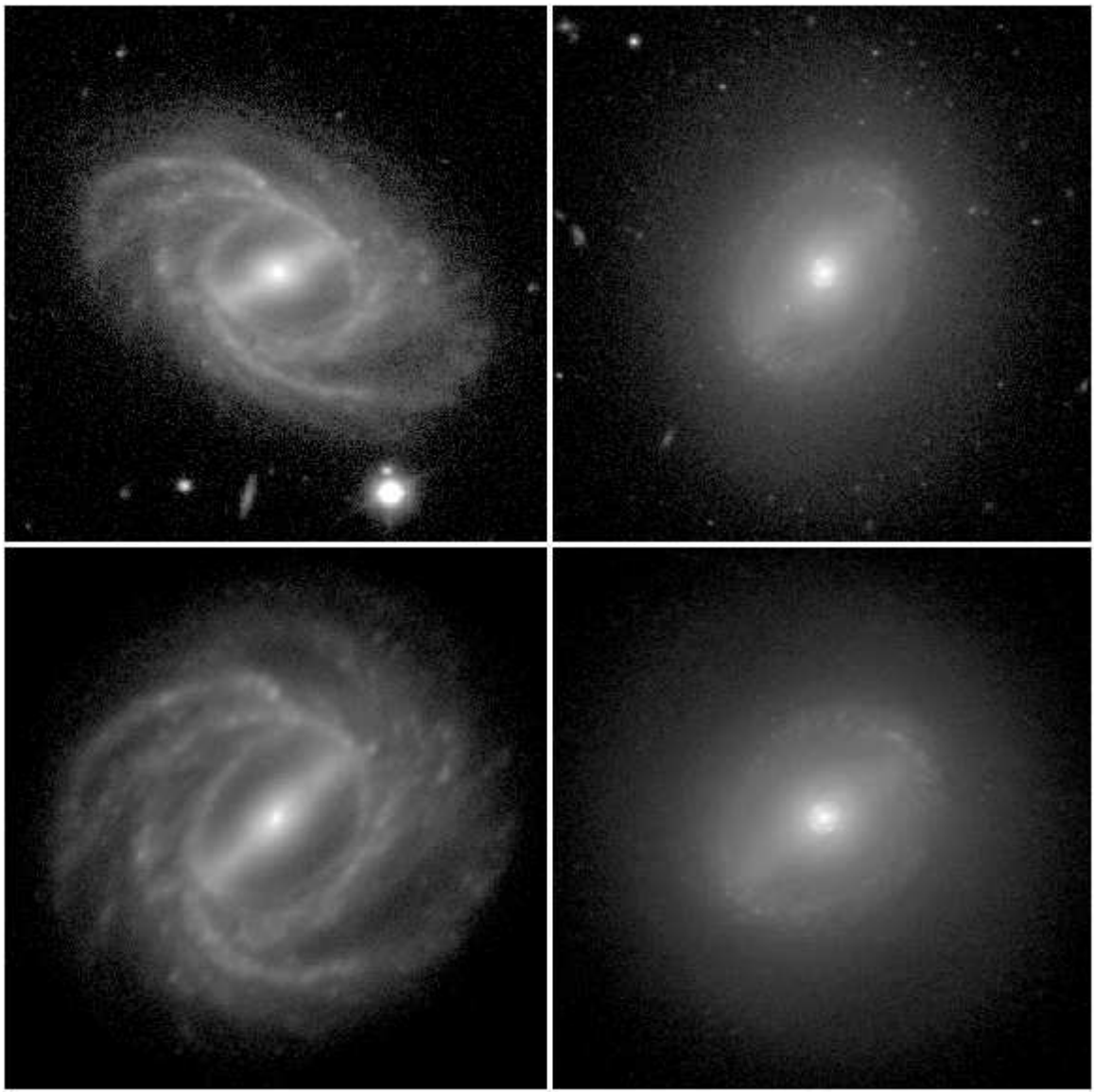}
\label{sample}
\caption{$B$-band images of NGC 2523 (left) and 4245 (right).
The upper images were obtained with the Nordic Optical Telescope in 2003
and 2004 (Laurikainen et al. 2005) and are presented as illustrated in the
de Vaucouleurs Atlas of Galaxies (Buta, Corwin, and Odewahn 2007). North is
at the top and east is to the left in each case. The lower images are 
deprojected to the disk plane and have the foreground stars removed. 
The images are 3.0$\arcmin$ by 3.0$\arcmin$.}
\end{figure}

\clearpage

\begin{figure}
\figurenum{2}
\includegraphics[angle=90,height=125mm]{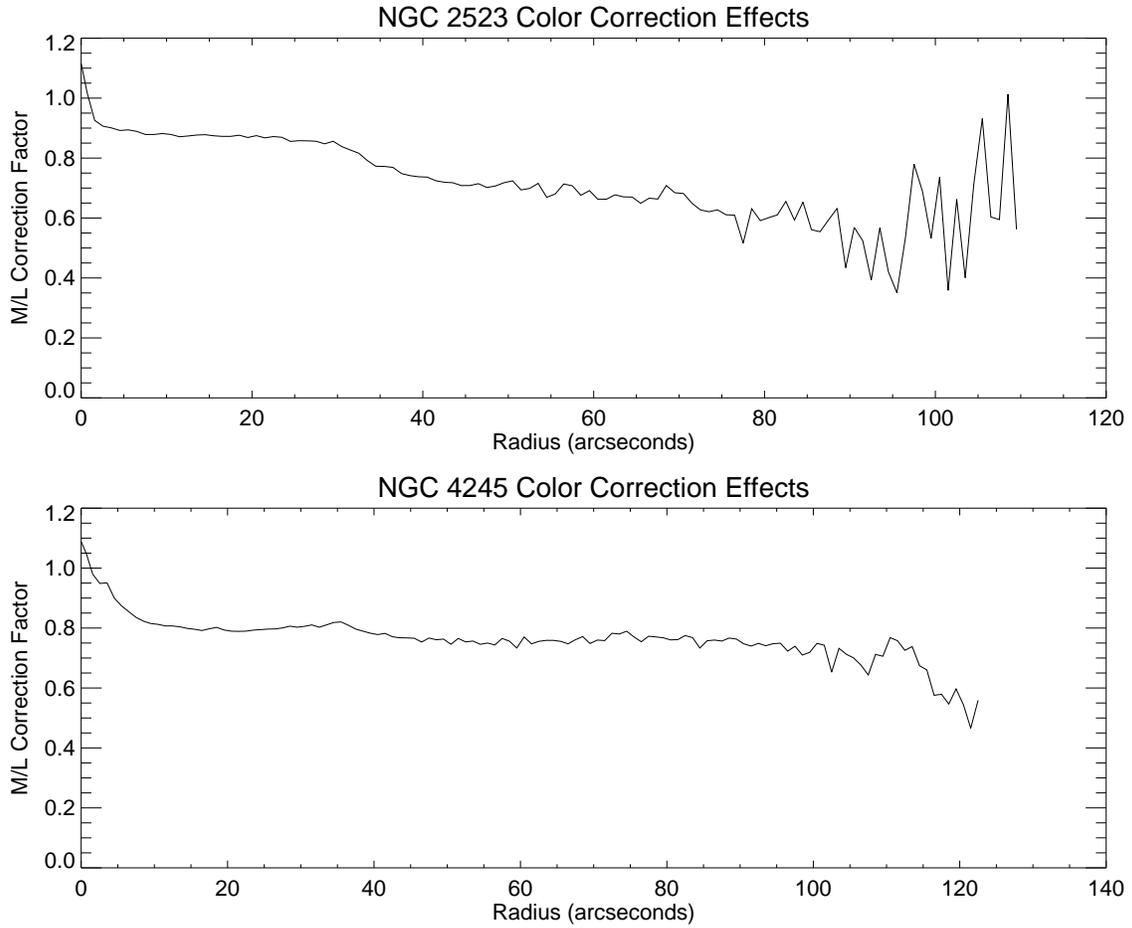}
\label{sample}
\caption{Plots of the radial M/L profiles derived
from the formation epoch with bursts model described in Bell \& de Jong (2001)
for NGC 2523 and NGC 4245.}
\end{figure}

\clearpage

\begin{figure}
\figurenum{3}
\includegraphics[angle=90,height=125mm]{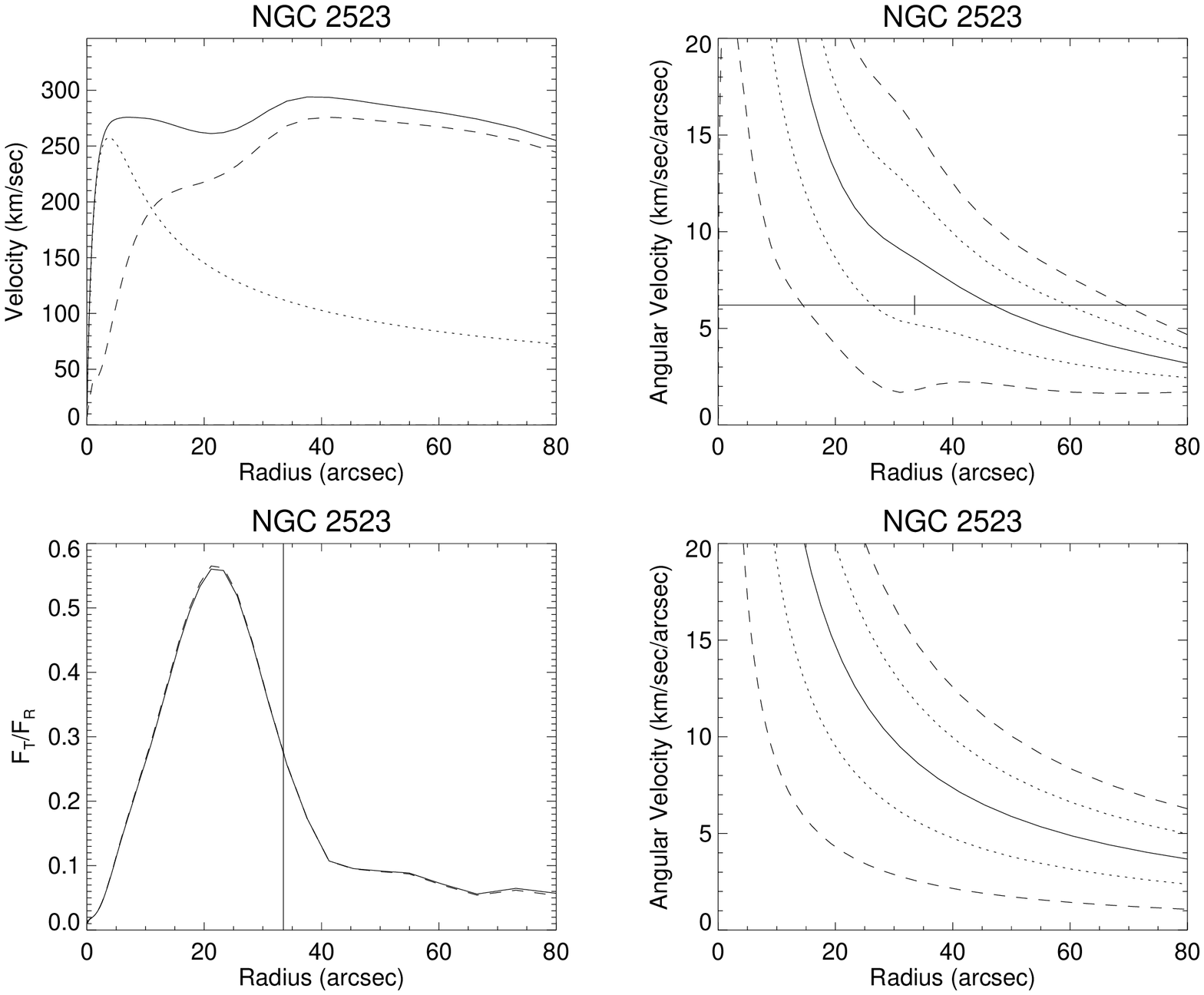}
\label{sample}
\caption{Plots of circular-speed curves, Lindblad precession frequency
curves, and the radial variation of the tangential force amplitude ($F_{T}$) 
normalized by the axisymmetric disk force ($F_{R}$) for NGC 2523. The 
upper left panel shows the rotation curve (solid curve) of the galaxy
derived from the mass model based on the M/L corrected $K_{s}$-band image. The
bulge (dotted curve) and disk components (dashed curve) are also shown.
The curves have been scaled so that the maximum
rotation velocity is equal to 294 km s$^{-1}$ (Treuthardt et al. 2007). The
panels on the right show the precession frequency curves derived from the mass
model rotation curve (upper right) and a constant rotation curve of 294 km
s$^{-1}$ (lower right). From left to right, the curves correspond to
$\Omega - \kappa/2$, $\Omega - \kappa/4$, $\Omega$, $\Omega + \kappa/4$,
$\Omega + \kappa/2$. $\Omega$ is the circular angular velocity and $\kappa$
is the epicyclic frequency. The horizontal line in the upper 
right panel indicates our best estimate of $\Omega_p$ = 6.2 km s$^{-1}$ 
arcsec$^{-1}$. The short vertical line corresponds to the error range in 
$\Omega_p$ and is placed at the estimated bar radius of 33.5$\arcsec$. The 
lower left panel shows the $F_{T}/F_{R}$
profile reaching a maximum of 0.56 at approximately 21$\arcsec$ (solid curve).
A dashed curve, which nearly coincides with the solid curve, shows the 
$F_{T}/F_{R}$ profile for the case of an added halo
component as shown in Figure 9 and discussed in section 3. The vertical line 
shows the estimated bar radius.}
\end{figure}

\clearpage

\begin{figure}
\figurenum{4}
\includegraphics[angle=90,height=125mm]{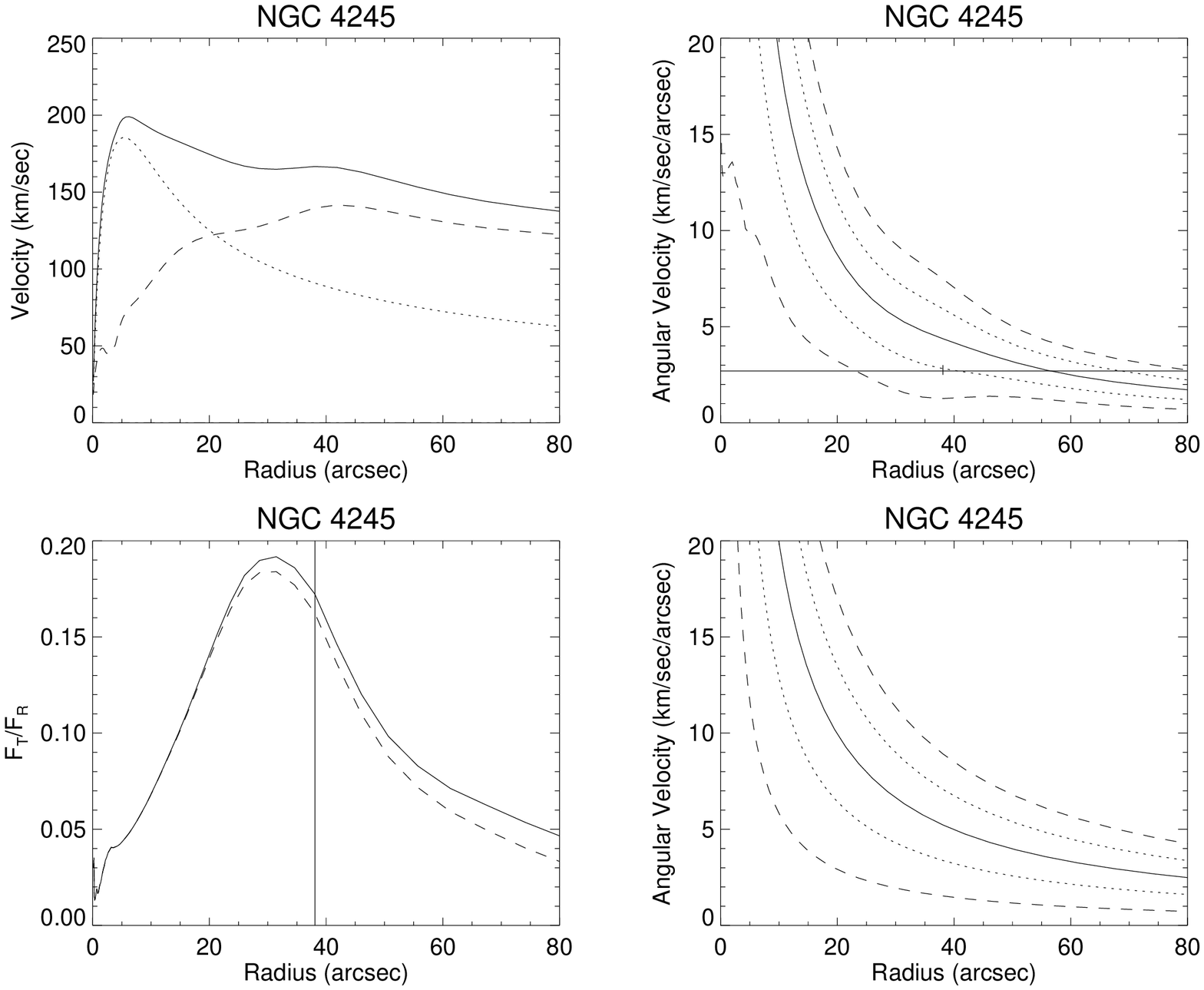}
\label{sample}
\caption{Plots of circular-speed curves, Lindblad precession frequency
curves, and the radial variation of the tangential force amplitude ($F_{T}$) 
normalized by the axisymmetric disk force ($F_{R}$) for NGC 4245. 
The upper left panel shows the rotation curve (solid curve) of the galaxy
derived from the mass model based on the M/L corrected $K_{s}$-band image. The
bulge (dotted curve) and disk components (dashed curve) are also shown.
The curves have been scaled so that the maximum
rotation velocity is equal to 199 km s$^{-1}$ (Treuthardt et al. 2007). The
panels on the right show the precession frequency curves derived from the mass
model rotation curve (upper right) and a constant rotation curve of 199 km
s$^{-1}$ (lower right). From left to right, the curves correspond to
$\Omega - \kappa/2$, $\Omega - \kappa/4$, $\Omega$, $\Omega + \kappa/4$,
$\Omega + \kappa/2$. $\Omega$ is the circular angular velocity and $\kappa$
is the epicyclic frequency. The horizontal line in the upper 
right panel indicates our best estimate of $\Omega_p$ = 2.7 km s$^{-1}$ 
arcsec$^{-1}$. The short vertical line corresponds to the error range in 
$\Omega_p$ and is placed at the estimated bar radius of 38.1$\arcsec$. The 
lower left panel 
shows the $F_{T}/F_{R}$ 
profile reaching a maximum of 0.19 at approximately 31$\arcsec$ (solid curve).
The dashed curve shows the $F_{T}/F_{R}$ profile for the case of an added halo
component as shown in Figure 9 and discussed in section 3. The vertical line 
shows the estimated bar radius.}
\end{figure}

\clearpage

\begin{figure}
\figurenum{5}
\includegraphics[angle=90,height=125mm]{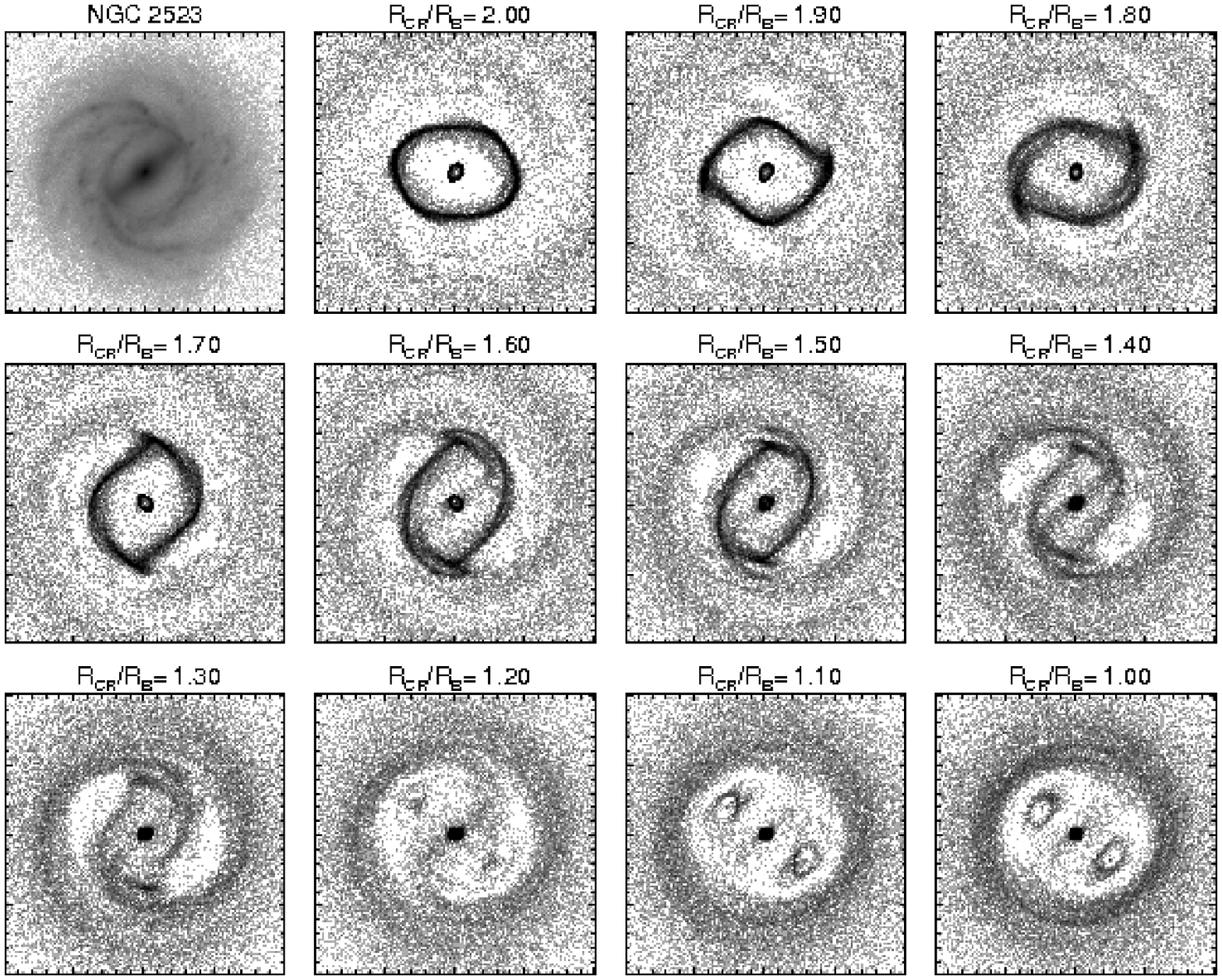}
\label{sample}
\caption{A $B$-band, disk-plane image of NGC 2523 (upper left panel) showing
the SB(r)b morphology. The remaining panels are 
gas simulations showing the disk-plane morphology at 7 bar rotations for 
different pattern speeds corresponding to the shown $R_{CR}/R_{B}$ values. 
The deprojected value of $R_{B}$ is 33.5$\arcsec$. Each panel
is 200$\arcsec$ by 200$\arcsec$.}
\end{figure}

\clearpage

\begin{figure}
\figurenum{6}
\includegraphics[angle=90,height=125mm]{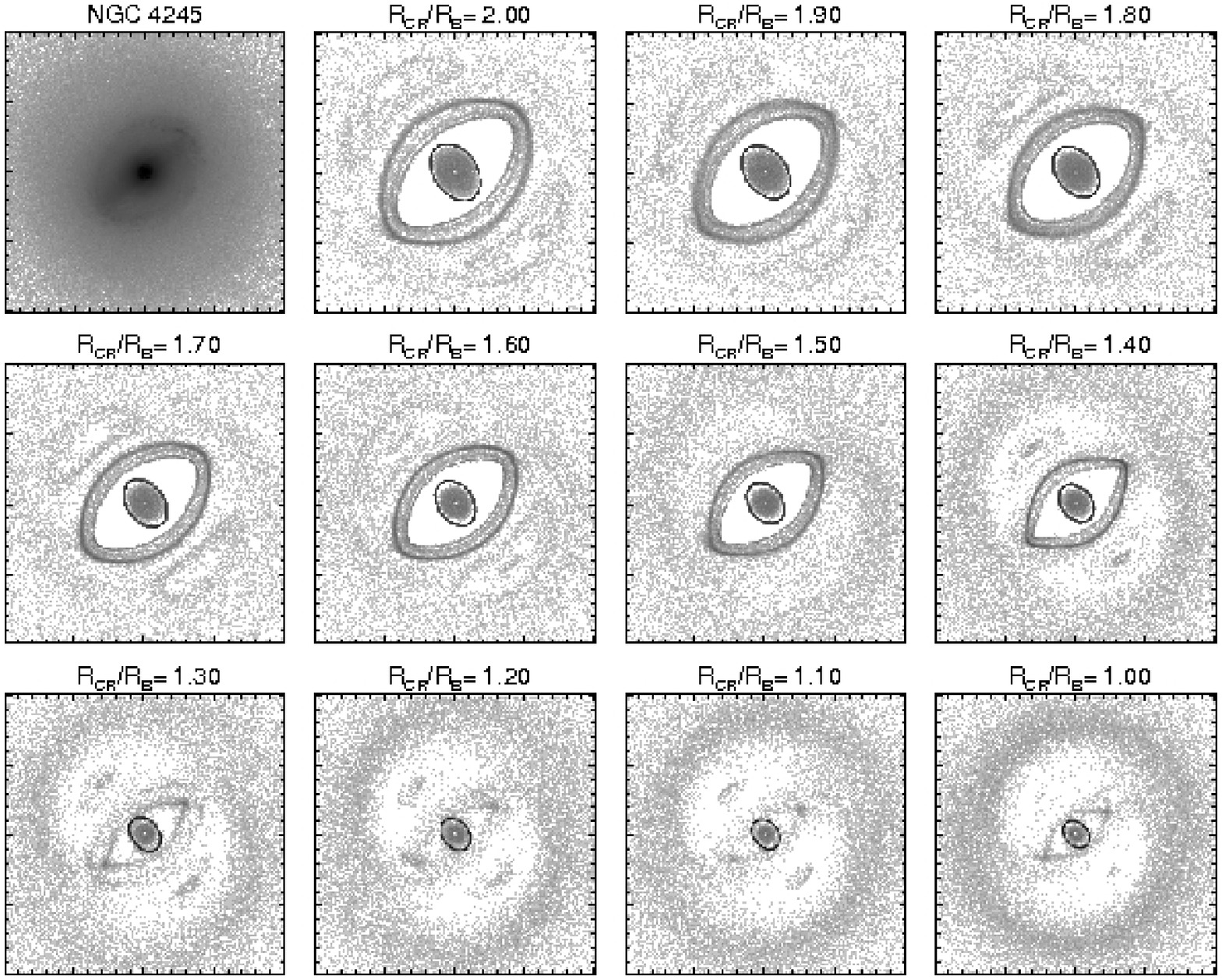}
\label{sample}
\caption{A $B$-band, disk-plane image of NGC 4245 (upper left panel) showing
the SB(r)0/a morphology. The remaining panels are
gas simulations showing the disk-plane morphology at 12 bar rotations for 
different pattern speeds corresponding to the shown $R_{CR}/R_{B}$ values. The 
deprojected value of $R_{B}$ is 38.1$\arcsec$. Each panel
is 200$\arcsec$ by 200$\arcsec$.}
\end{figure}

\clearpage

\begin{figure}
\figurenum{7}
\includegraphics[angle=90,trim=50 0 75 0,clip=true,scale=0.5]{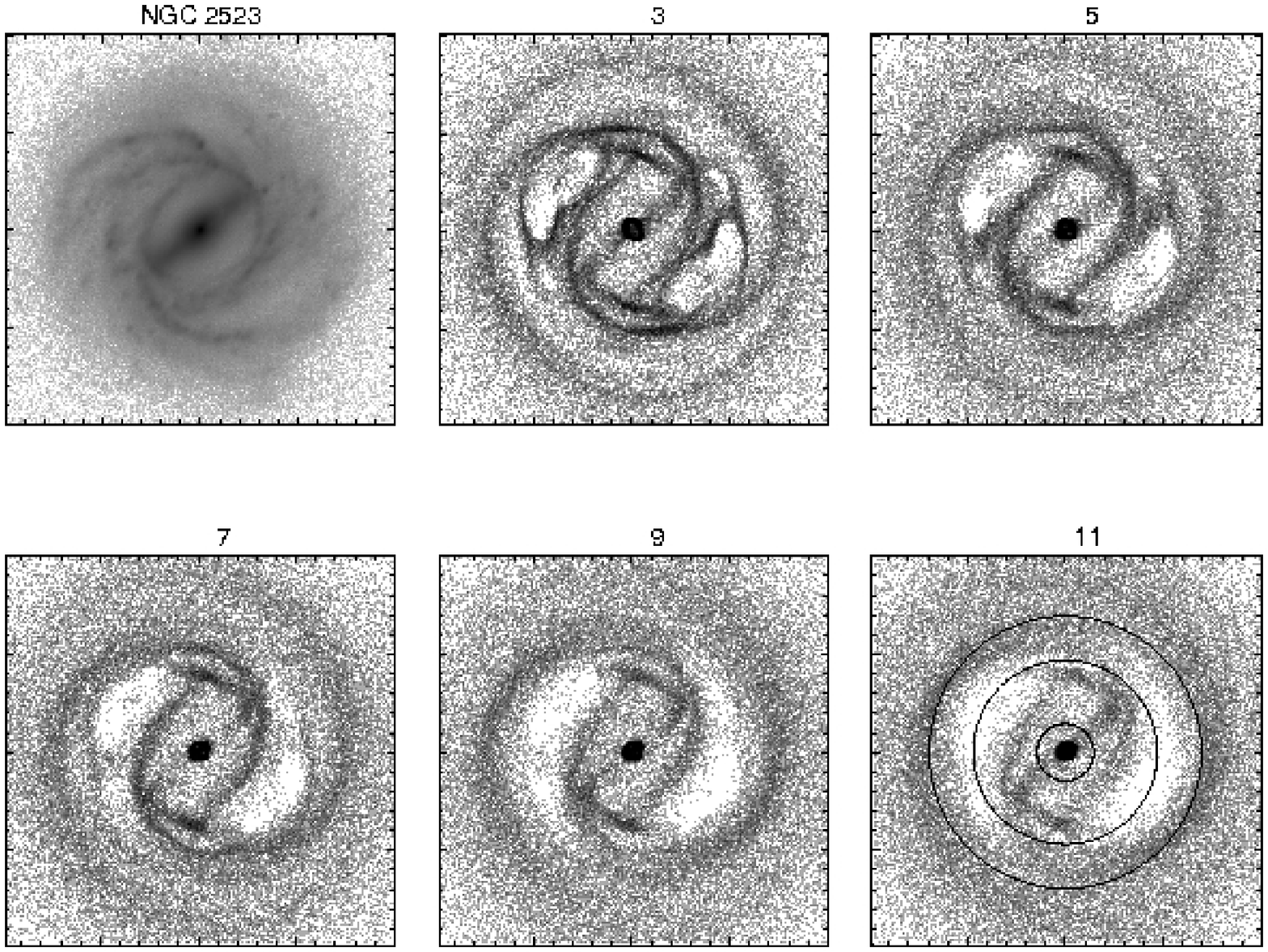}
\includegraphics[angle=90,trim=50 0 75 0,clip=true,scale=0.5]{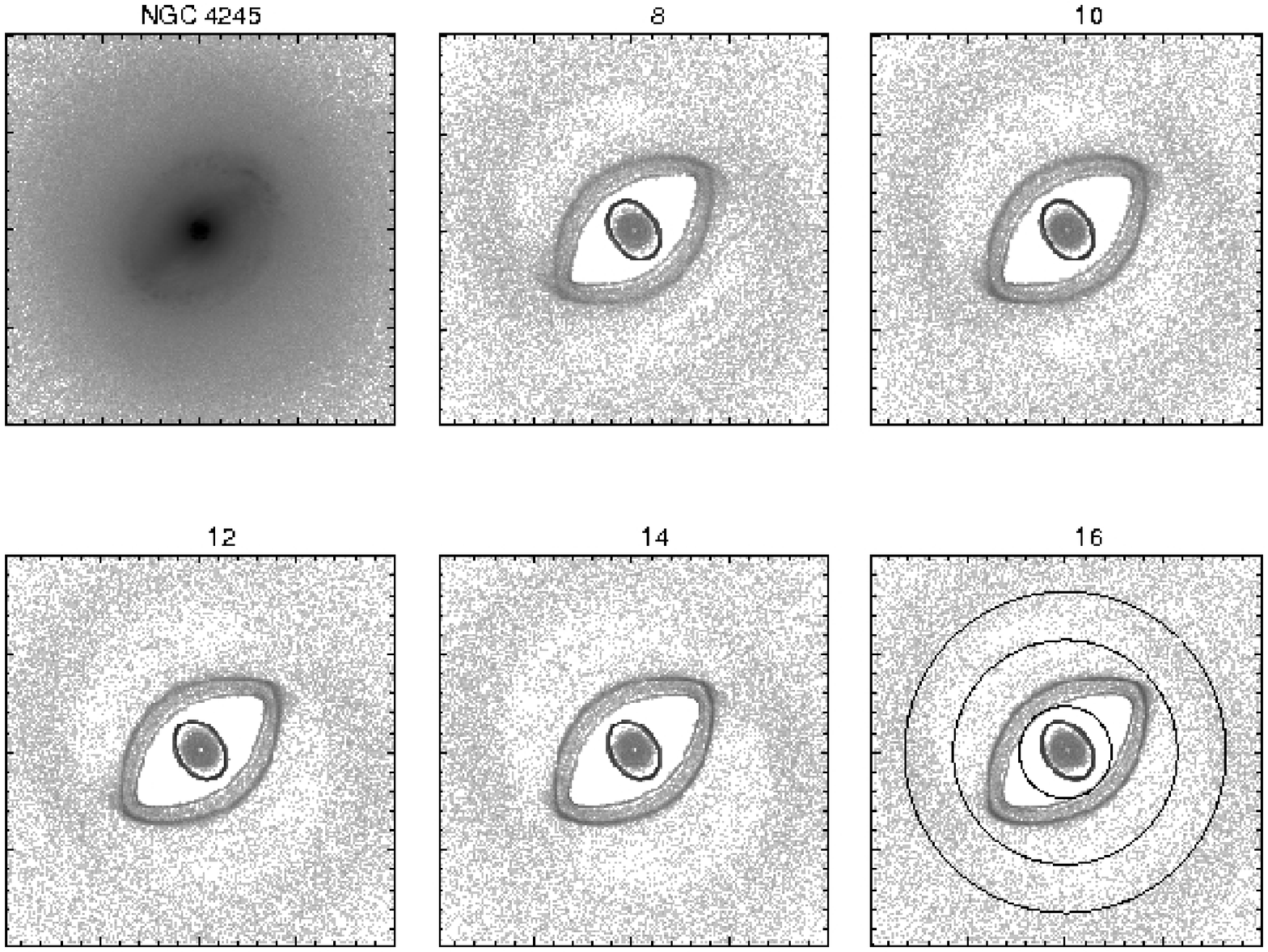}
\label{sample}
\caption{Two sets of six panels comparing the $B$-band and gas 
simulation morphology of NGC 2523 (upper set) and NGC 4245 (lower set) in the
disk-plane. For NGC 2523, the upper left panel shows the $B$-band SB(r)b 
morphology. The remaining panels are
gas simulations showing the morphology from 3 to 11 bar rotations for a
pattern speed corresponding to $R_{CR}/R_{B}$ of 1.4. For NGC 4245, the upper 
left panel shows the $B$-band SB(r)0/a morphology. The remaining panels are
gas simulations showing the morphology from 8 to 16 bar rotations for a
pattern speed corresponding to $R_{CR}/R_{B}$ of 1.5. The lower right panel 
in each set has circles overlayed which indicate (from smallest to largest 
radius) the locations of the inner Lindblad resonance, corotation, 
and the outer Lindblad resonance. Each panel is 200$\arcsec$ by 200$\arcsec$.}
\end{figure}

\clearpage

\begin{figure}
\figurenum{8}
\includegraphics[height=125mm,trim=0 0 260 0,clip=true,angle=90,scale=1.2]{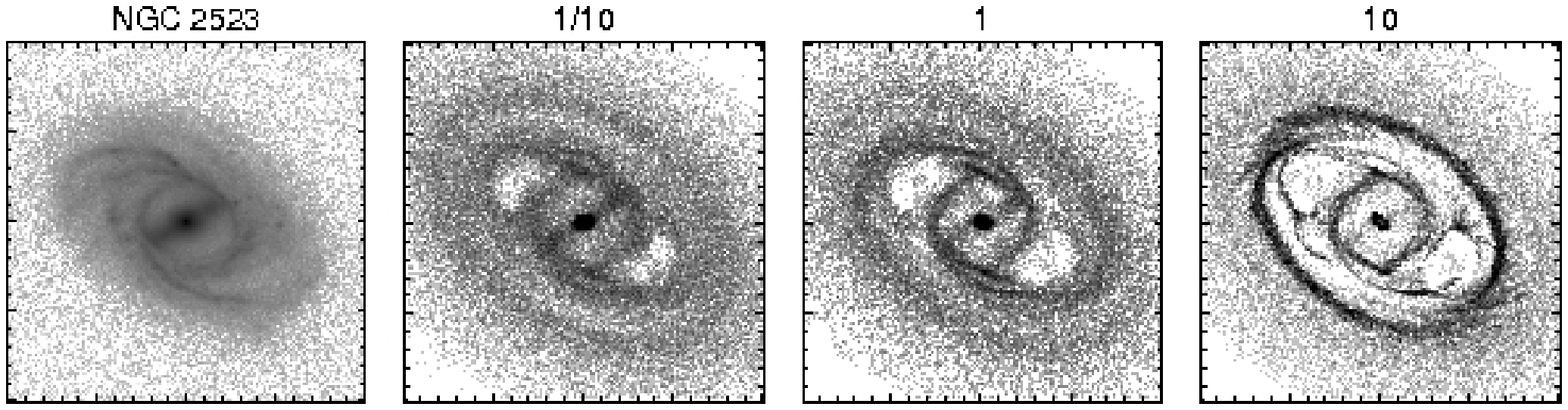}
\includegraphics[height=125mm,trim=300 0 0 0,clip=true,angle=90,scale=1.2]{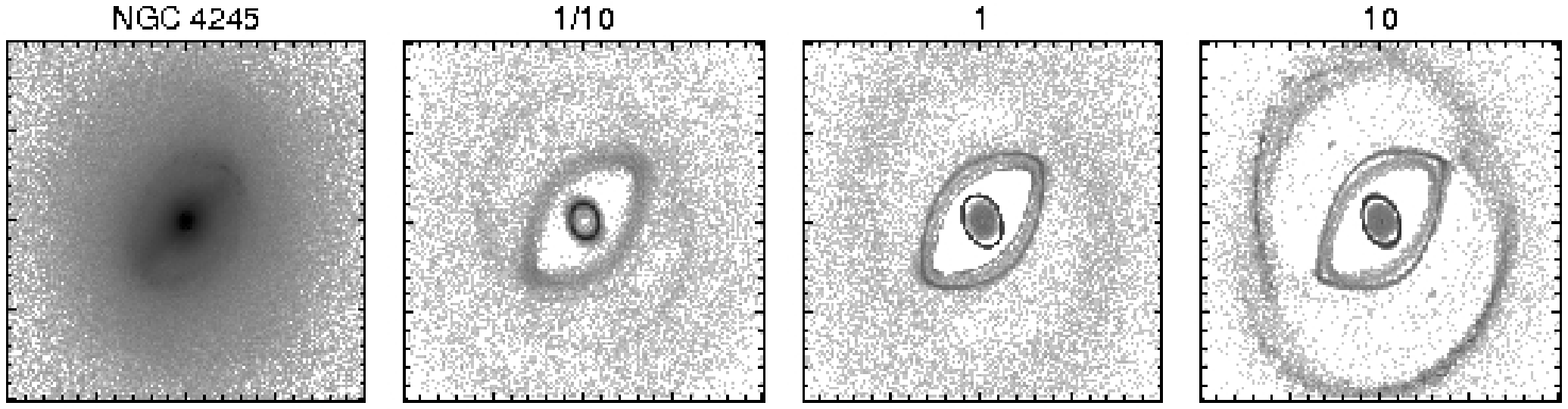}
\label{sample}
\caption{Examples of the difference in morphology of the simulated galaxies 
when
using different particle sizes. The left-most panels show a $B$-band image of
NGC 2523 (upper) and NGC 4245 (lower). The remaining panels show the 
simulated gas particle morphology of
the galaxy at 7 bar rotations and $R_{CR}/R_{B}$ equal to 1.4 for NGC 2523 and
12 bar rotations and $R_{CR}/R_{B}$ equal to 1.5 for NGC 4245. The particle
radii used were, from left to right, 1/10, 1, and 10 times our standard of
1$\times$10$^{-2}$ arcseconds.}
\end{figure}

\clearpage

\begin{figure}
\figurenum{9}
\includegraphics[angle=90,height=63mm]{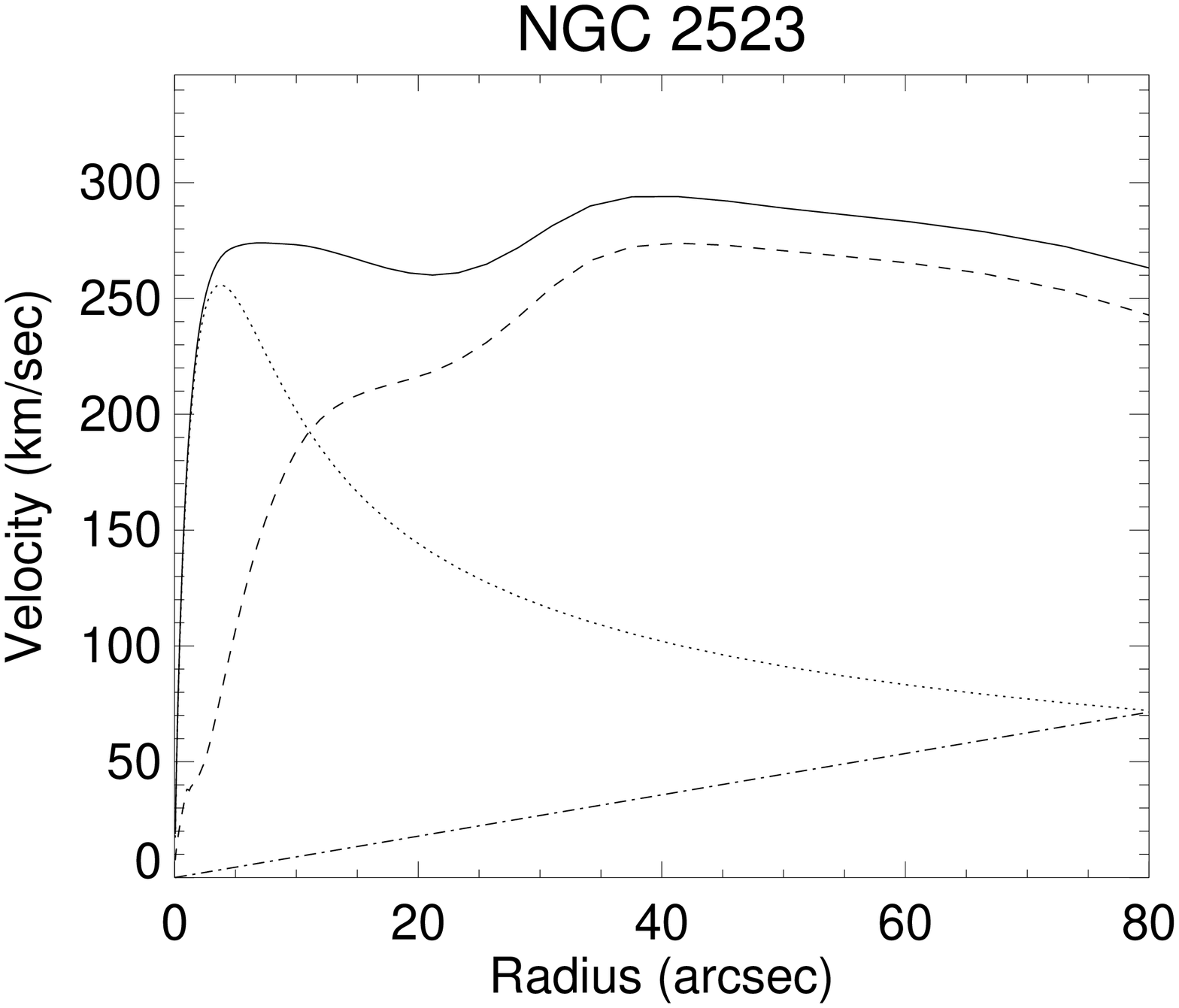}
\includegraphics[angle=90,height=63mm]{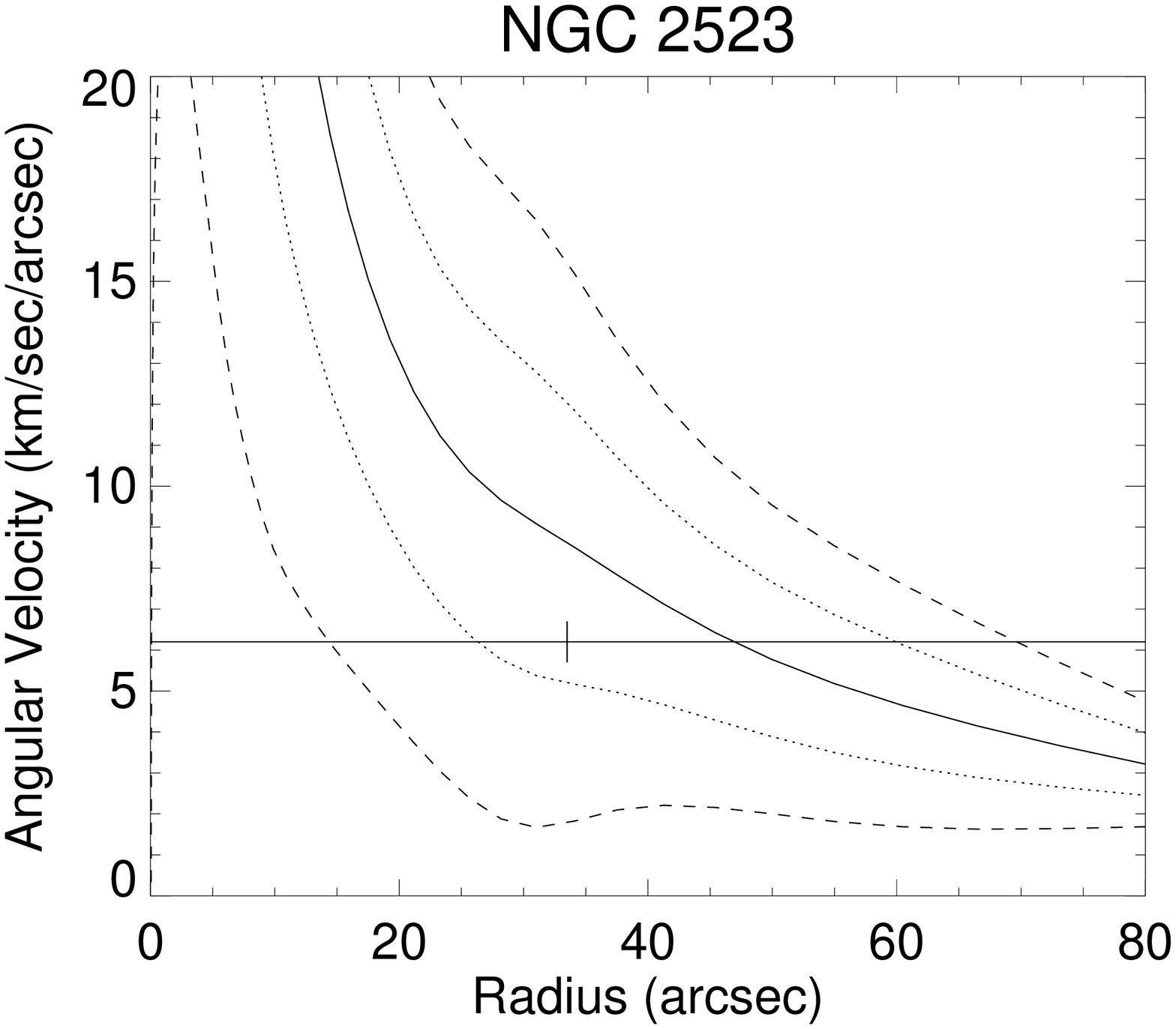}
\includegraphics[angle=90,height=63mm]{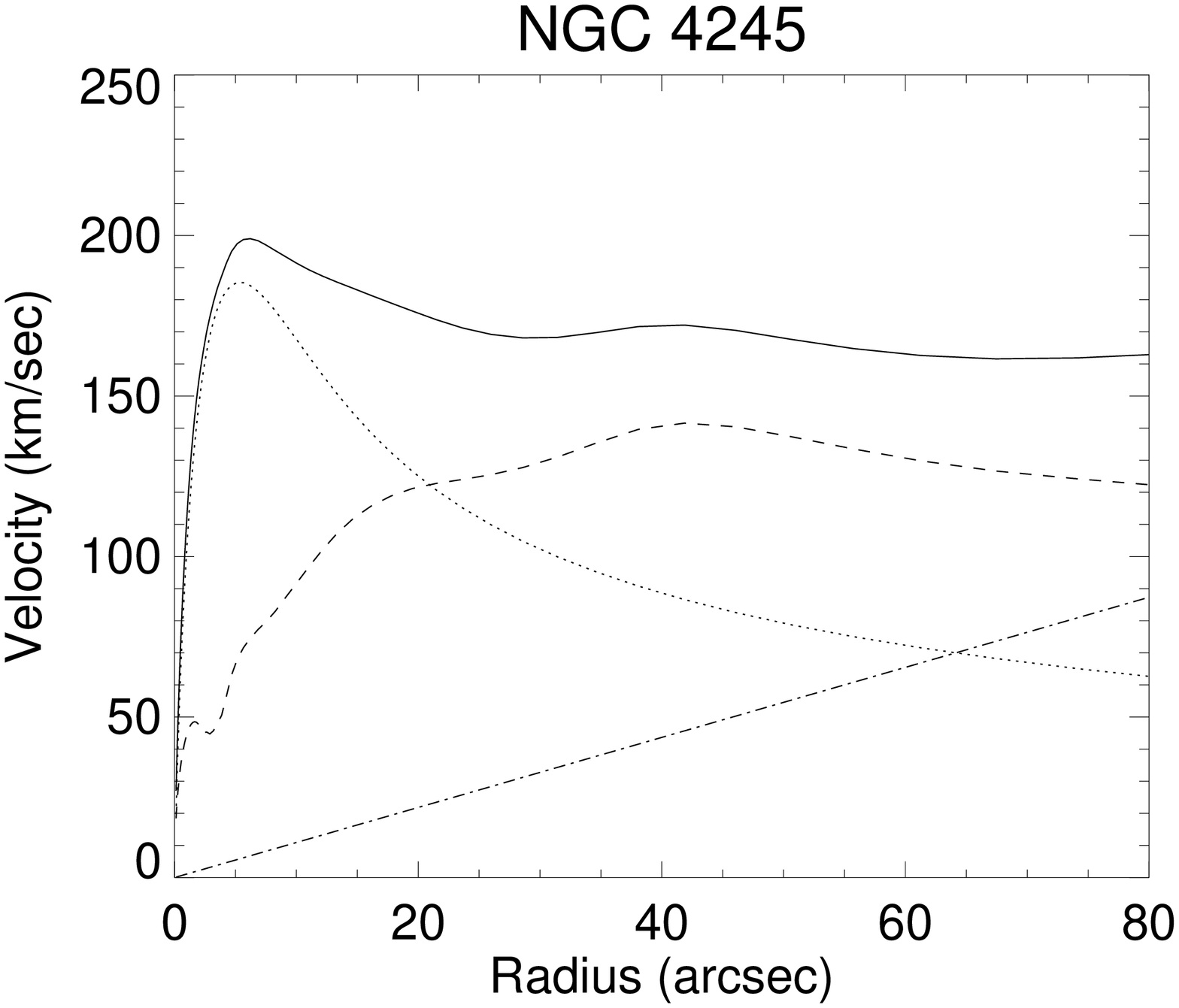}
\includegraphics[angle=90,height=63mm]{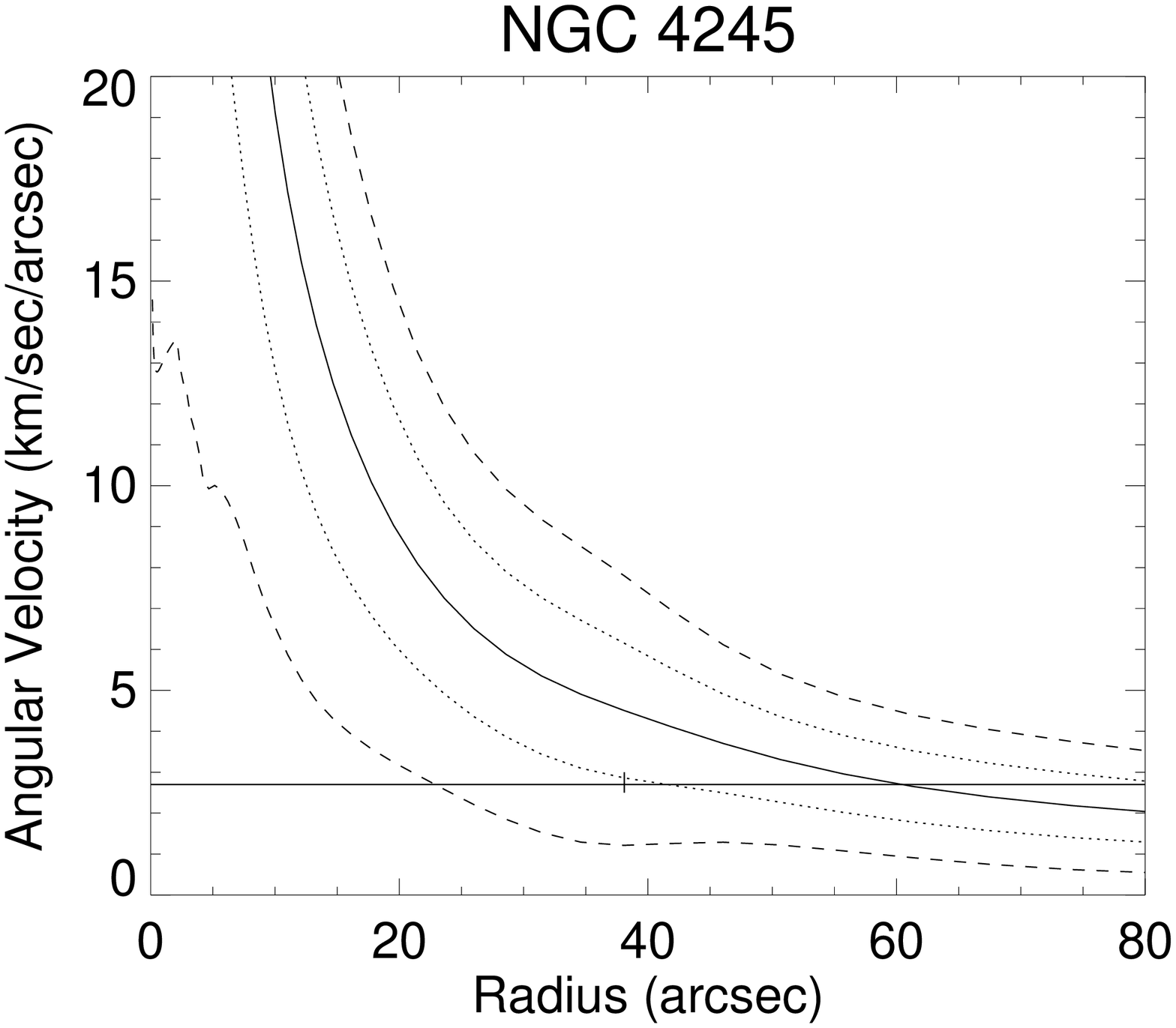}
\label{sample}
\caption{The plots shown here are similar to the upper plots in Figures 3 and
4, but with the addition of a halo component. The halo contribution to
the total circular-speed curve is shown as the dot-dashed curve in the
plots on the left. The corresponding Lindblad precession frequency curves are
shown on the right.}
\end{figure}

\clearpage

\begin{figure}
\figurenum{10}
\includegraphics[angle=90,trim=50 0 75 0,clip=true,scale=0.5]{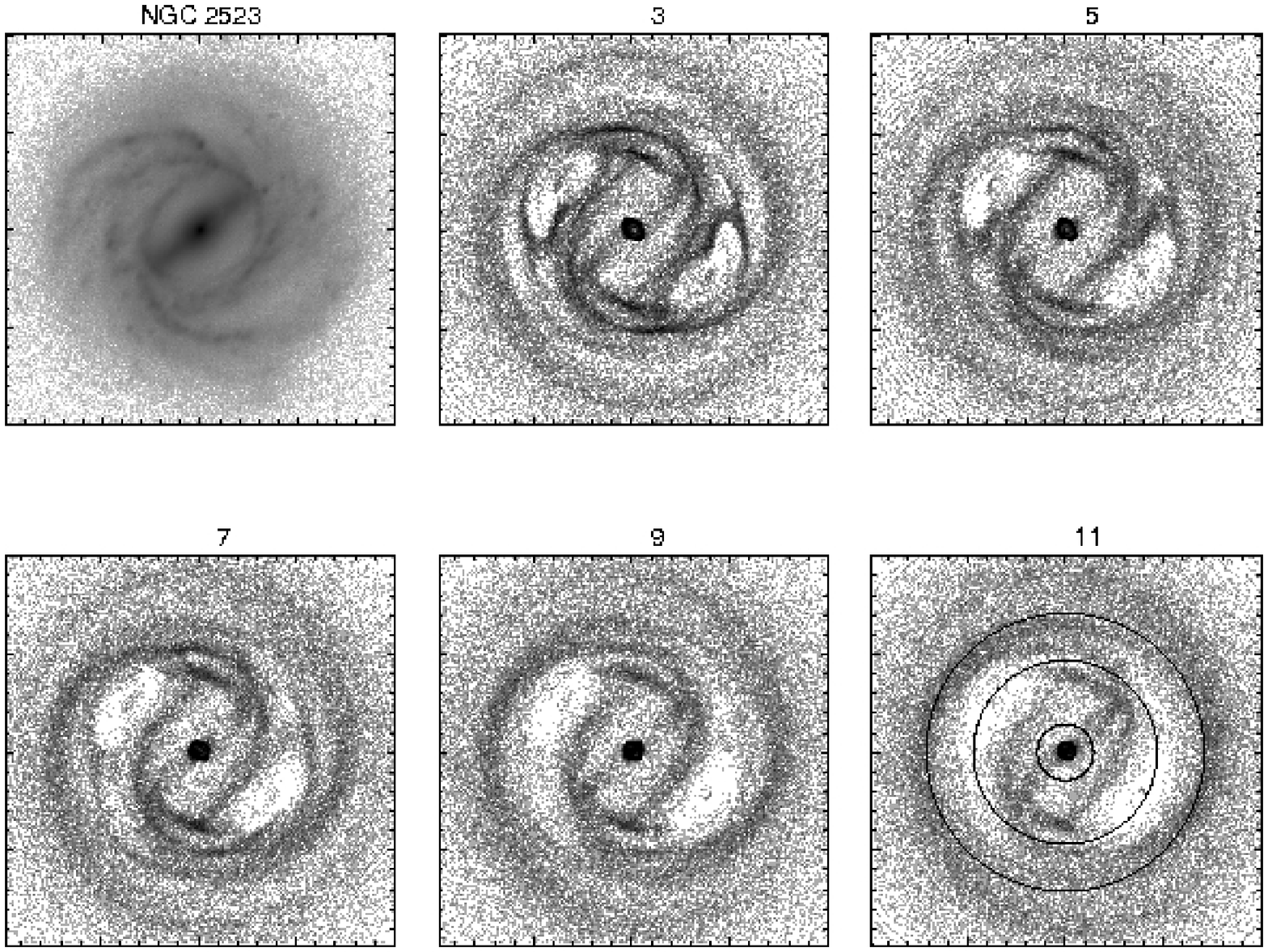}
\includegraphics[angle=90,trim=50 0 75 0,clip=true,scale=0.5]{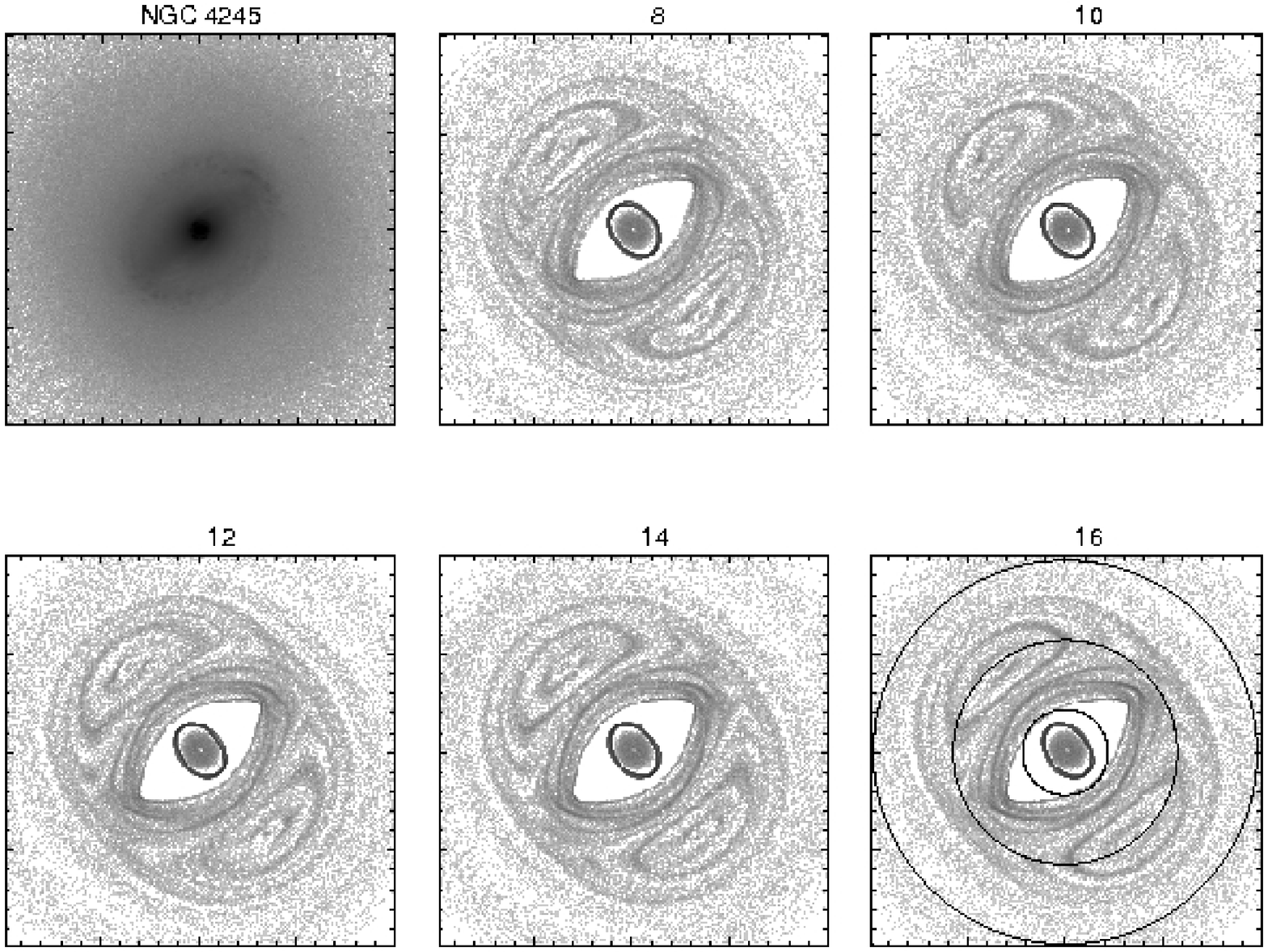}
\label{sample}
\caption{Plots similar to those shown in Figure 7 only with the contribution of
a halo component (shown in Figure 9) taken into account.}
\end{figure}

\clearpage

\begin{figure}
\figurenum{11}
\includegraphics[height=130mm,trim=30 0 75 0,clip=true,angle=90,scale=0.9]{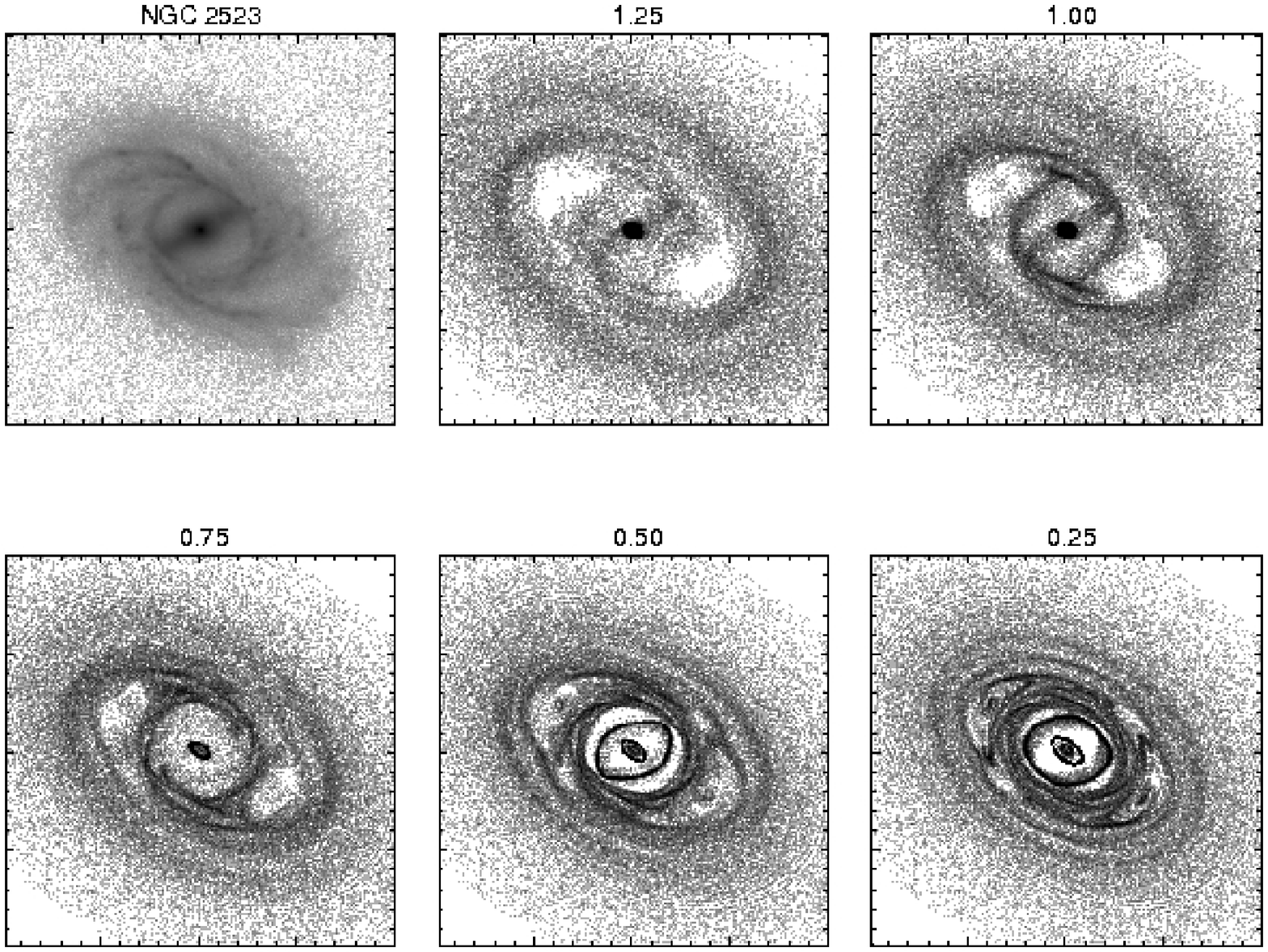}
\includegraphics[height=130mm,trim=50 0 75 0,clip=true,angle=90,scale=0.9]{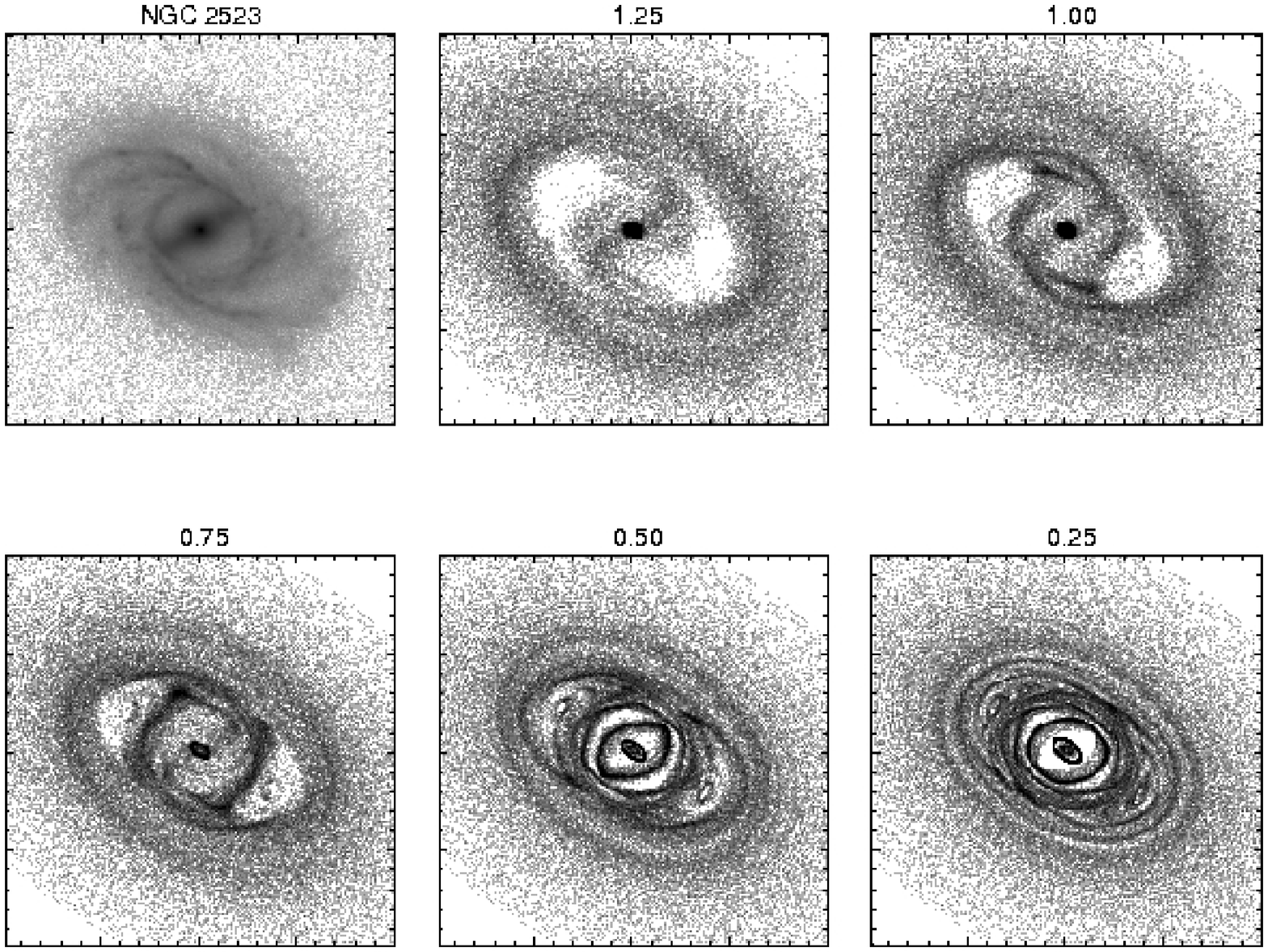}
\label{sample}
\caption{Images showing the morphological effects of applying different bar
amplitudes to models of NGC 2523. The top two rows compare the $B$-band image
(upper left) to our best model ($\Omega_p = 6.2$ km s$^{-1}$ arcsec$^{-1}$ and
7 bar rotations) with the bar amplitude varying from 1.25 to 0.25. The bottom
two rows compare the $B$-band image to models using the average pattern speed
found by Treuthardt et al. (2007; $\Omega_p = 6.6$ km s$^{-1}$ arcsec$^{-1}$
and 7 bar rotations) with the bar amplitude also varying from 1.25 to 0.25.}
\end{figure}

\clearpage

\begin{figure}
\figurenum{12}
\includegraphics[height=130mm,trim=30 0 75 0,clip=true,angle=90,scale=0.9]{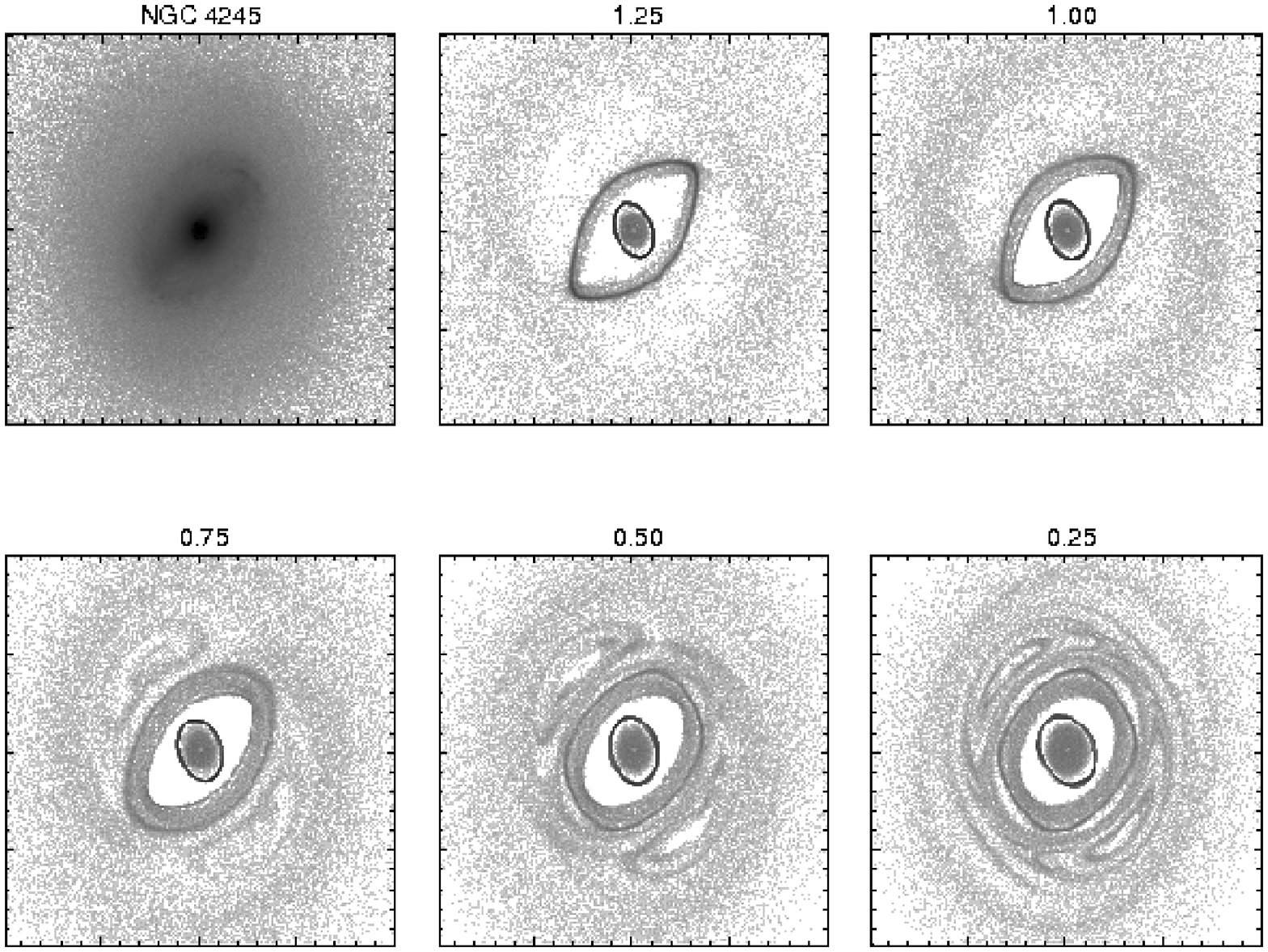}
\includegraphics[height=130mm,trim=50 0 75 0,clip=true,angle=90,scale=0.9]{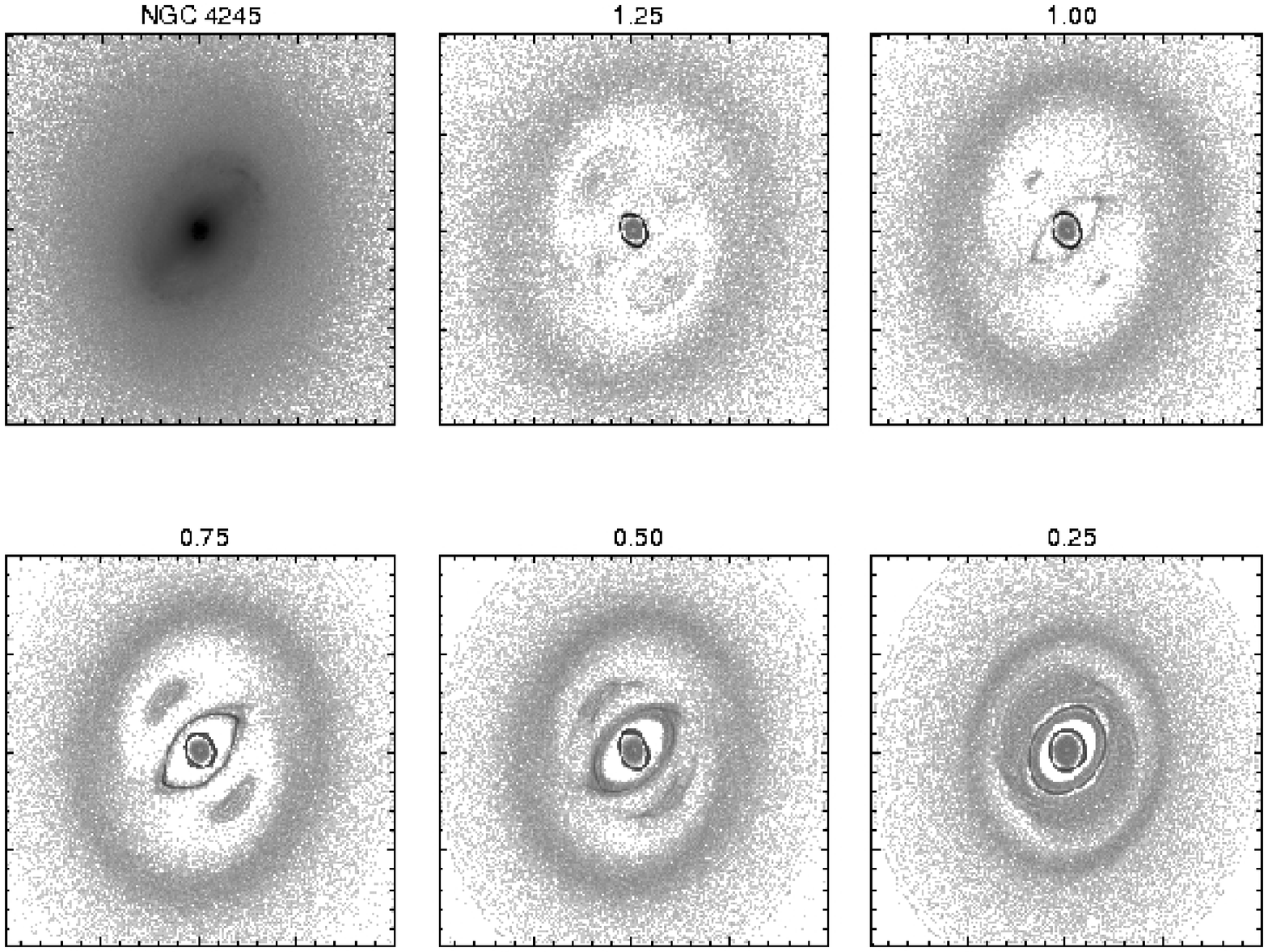}
\label{sample}
\caption{Images showing the morphological effects of applying different bar
amplitudes to models of NGC 4245. The top two rows compare the $B$-band image
(upper left) to our best model ($\Omega_p = 2.7$ km s$^{-1}$ arcsec$^{-1}$ and
12 bar rotations) with the bar amplitude varying from 1.25 to 0.25. The bottom
two rows compare the $B$-band image to models using the average pattern speed
found by Treuthardt et al. (2007; $\Omega_p = 4.7$ km s$^{-1}$ arcsec$^{-1}$
and 12 bar rotations) with the bar amplitude also varying from 1.25 to 0.25. }
\end{figure}

\clearpage

\begin{figure}
\figurenum{13}
\includegraphics[angle=90,trim=0 0 75 0,clip=true,scale=0.5]{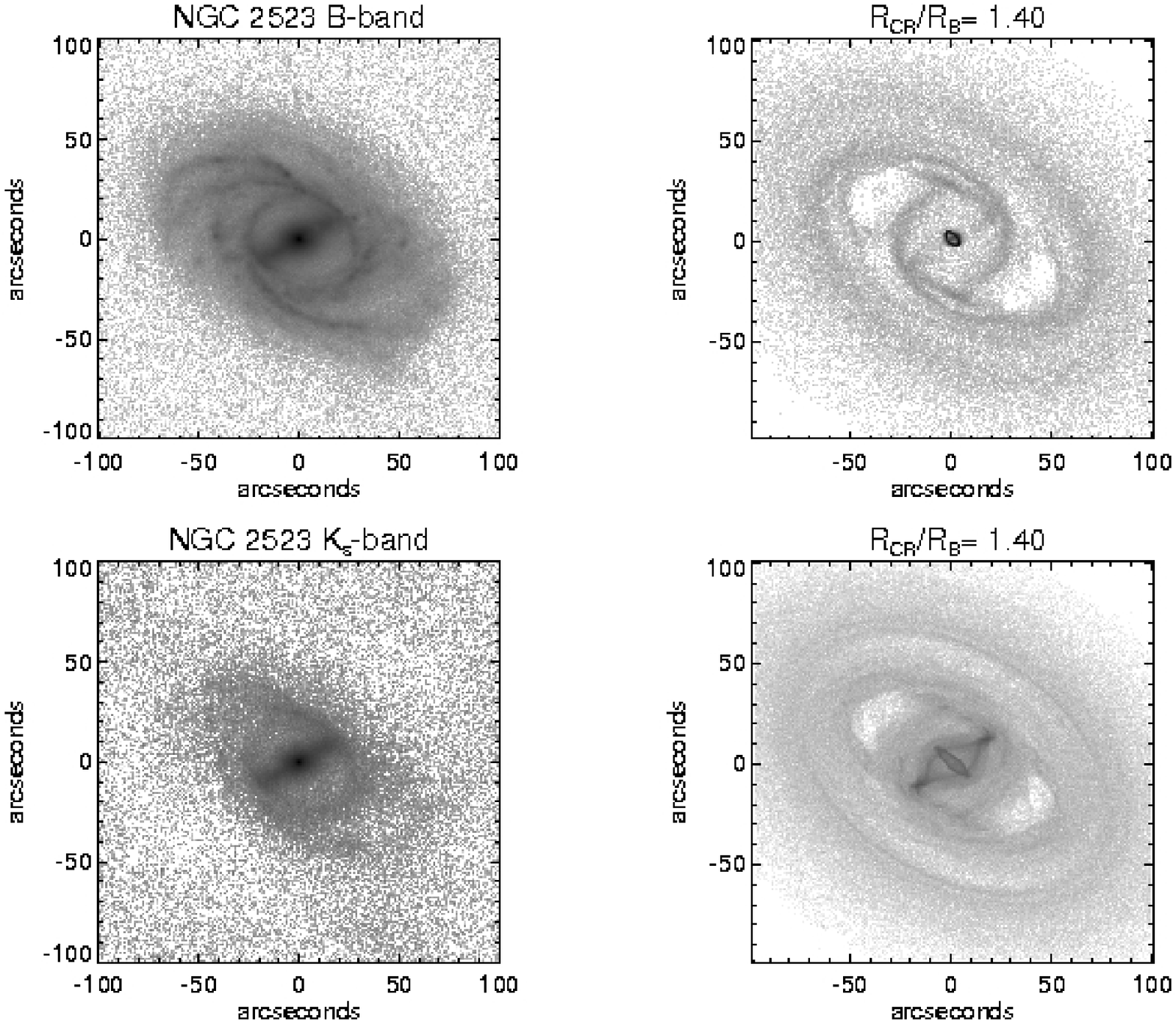}
\newline
\includegraphics[angle=90,trim=50 0 75 0,clip=true,scale=0.5]{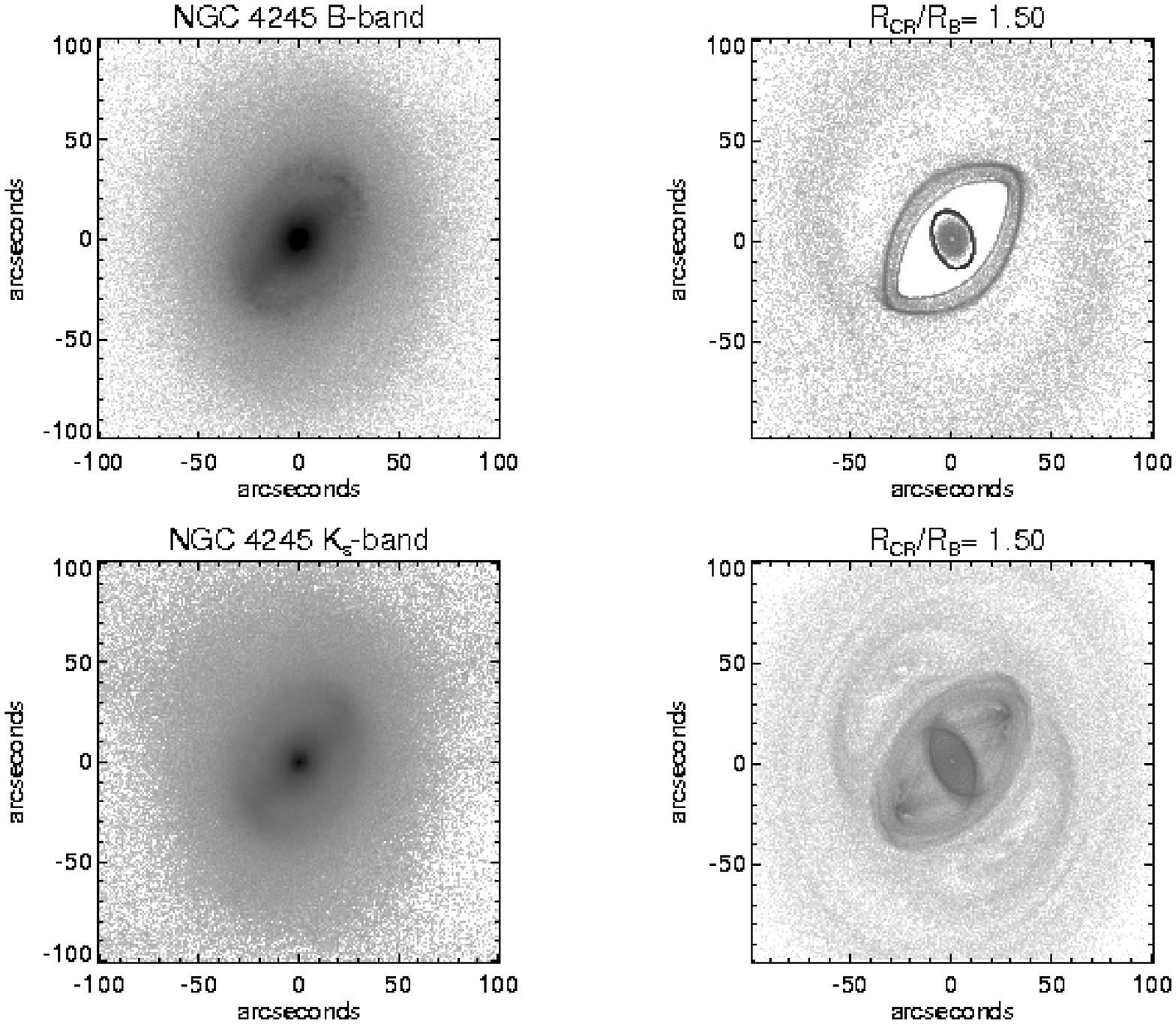}
\label{sample}
\caption{Images showing the observed and simulated morphologies of NGC 2523 
(upper four plots) and NGC 4245 (lower four plots). The $B$-band (upper left),
simulated gas (upper right), $K_s$-band (lower left), and simulated stellar 
morphology (lower right) of each galaxy is displayed. The simulated gas and 
stellar 
morphology of NGC 2523 is shown at 7 bar rotations with $R_{CR}/R_{B}$ = 1.4 
and no additional halo. The simulated gas and stellar morphology of NGC 4245
is shown at 12 bar rotations with $R_{CR}/R_{B}$ = 1.5 and no additional halo.
The simulated morphologies correspond to our best simulations.
The initial radial velocity dispersion of both the gas and stellar test 
particles is 10\% of the circular velocity. The stellar test particles are 
non-colliding, unlike the gas particles.}
\end{figure}

\end{document}